\documentclass[aps,prd,amsmath,amssymb,superscriptaddress,showpacs]{revtex4}
\usepackage{epsfig,graphics,color}
\usepackage{graphicx}
\usepackage{dcolumn}
\usepackage{bm}
\newcommand \etc {{\it etc.} }
\newcommand \ie  {{\it i.e.} }

\newcommand \f {\not\!}

\newcommand \nf {\tilde{n}}

\newcommand \pk {E_{p-k}}

\newcommand \kd  {\delta}

\newcommand \w  {\omega}
\newcommand \fw {{\bf w}}
\newcommand \fp {{\bf p}}

\newcommand \fk {{\bf k}}
\newcommand \fq {{\bf q}}

\newcommand \h {\theta}

\newcommand \vk {\vec{k}}
\newcommand \vq {\vec{q}}

\newcommand \vx {\vec{x}}
\newcommand \vp {\vec{p}}

\newcommand \mat {{\mathcal M}}
\newcommand \mps {\mu}
\newcommand \mas {\bar{\mu}}
\newcommand \g {\gamma}
\newcommand \ro {\rho}

\newcommand \e {\epsilon}
\newcommand \ve {\varepsilon}

\newcommand \A {\alpha}
\newcommand \B {\beta}

\newcommand \lc {\langle}
\newcommand \rc {\rangle}
\newcommand \ra {\rightarrow}

\newcommand \D {\Delta}
\newcommand \sg {\sigma}

\newcommand \nt {\noindent}

\newcommand \Ow {\Omega}
\newcommand \dsc {\mbox{Disc}}
\newcommand \Ro {\mathcal{R}}
\newcommand \ad {a^{\dag}}

\newcommand \ua {\uparrow}
\newcommand \da {\downarrow}
\newcommand {\llb} { \left[ \frac{\mbox{}}{\mbox{}} \right.}
\newcommand {\lrb} { \left. \frac{\mbox{}}{\mbox{}} \right] }
\newcommand \mt {\mathcal{T}}
\newcommand \mb {\mathcal{B}}
\newcommand \gmn {g^{\mu \nu}}
\newcommand \gmr {g^{\mu \ro}}
\newcommand \gnr {g^{\nu \ro}}
\newcommand \af {\frac{1}{2}}
\newcommand \ini {\infty}
\newcommand \res {\mbox{Res.}}
\newcommand \kD {\Delta}
\newcommand \ma {\mathcal{A}}
\newcommand \bvec{\left( \begin{array}{c} }
\newcommand \evec{\end{array} \right)}

\newcommand \eg {{\it e.g.}}  
\newcommand \bea{\begin{eqnarray} }
\newcommand \eea{\end{eqnarray} }
\newcommand \nn {\nonumber}
\newcommand {\be} {\begin{equation}}
\newcommand {\ee} {\end{equation}}

\begin{document}

\title{ Broken symmetries and dilepton production from gluon fusion \\
 in a quark gluon plasma }

\author{A. Majumder}
\affiliation{Nuclear Science Division, 
Lawrence Berkeley National Laboratory\\
1 Cyclotron road, Berkeley, CA 94720}

\author{A. Bourque}
\affiliation{Physics Department, McGill University\\ 
3600 University Street, Montreal, QC, 
Canada H3A 2T8}
\author{C. Gale}
\affiliation{Physics Department, McGill University\\  
3600 University Street, Montreal, QC, 
Canada H3A 2T8}

\date{ \today}

\begin{abstract}
The observational consequences of certain 
broken symmetries in a thermalised quark gluon plasma are elucidated. 
The signature under study is the spectrum of dileptons 
radiating from the plasma, through gluon fusion. 
Being a pure medium effect, this channel is non-vanishing only in plasmas 
with explicitly 
broken charge conjugation invariance. The emission rates are also 
sensitive to rotational invariance through the constraints 
imposed by Yang's theorem. This theorem is interpreted in the 
medium via the destructive interference between various 
multiple scattering diagrams obtained in the spectator picture.
Rates from the fusion process are  
presented in comparison with those from the Born term. 
\end{abstract}


\preprint{LBNL-52689}

\maketitle


\section{INTRODUCTION}


Experiments are now underway at the Relativistic Heavy Ion Collider (RHIC) at 
Brookhaven National Laboratory to study nuclear collisions at very high energies. 
The aim is to create energy densities high enough for the production of a
state of essentially deconfined quarks and gluons: the quark gluon plasma (QGP). 
The QGP is rather short lived and soon hadronizes into a plethora of 
mesons and baryons. Hence, the existence of such a state in the 
history of a given 
collision must likely be surmised through a variety of indirect probes. 
One of the most promising signatures has been that of the electromagnetic 
probes: the spectrum of lepton pairs and real photons emanating from a 
given collision. These particles once produced interact only 
electromagnetically 
with the plasma. As a result they escape the plasma with almost no 
further rescattering and convey information from all time sectors of the
collision.

In this article, the focus will be on the spectrum 
of dileptons radiating from a heavy-ion collision. 
The primary motivation for measuring such a spectrum is 
the hope that the formation of a QGP in the history of a collision 
will produce a qualitative or quantitative difference in the observed rates. 
The measured quantity is the number of dileptons, usually 
binned according to their 
invariant mass. It is assumed that this may be estimated by the following 
factorized form,  

\begin{equation}
\frac{ dN_{e^+e^-} }{ dM } = \int_{\tau_0}^{t_f} dt \int_{x_-(t)}^{x_+(t)} 
\int_{z_-(t)}^{z_+(t)} \int_{y_-(t)}^{y_+(t)} d^3 x \int d^3 q
\frac{M}{q^0} \frac{d^4 R_{e^+ e^-}} {d^4 q} (q^0,\vq,T(t,\vx),\mu(t,\vx) ). 
\label{totaldilep}
\end{equation}

\nt where, $\frac{d^4 R_{e^+ e^-}}{d^4 q}$ is the number of lepton 
pairs produced per unit space
time, per unit four-momentum, from the unit cell at $(\vec{x},t)$ in a 
plasma in local equilibrium (local equilibrium is assumed here). 
Ostensibly, this depends on the 
four-momentum of the virtual 
photon ($q^0,\vq$), the temperature ($T$), and the relevant chemical potential ($\mu$). 
The temperature and relevant chemical potential are, in general, local properties 
for an expanding plasma and vary from point to point in the plasma as indicated.
Finally, the rates from each space time cell have to be integrated over the entire 
space time evolution of the plasma; where, 
the spatial limits of the expanding
plasma are represented by the variables 
$x_-(t),x_+(t)$, $y_-(t),y_+(t)$, $z_-(t),z_+(t)$.  
  
Many calculations of the dilepton radiation in the deconfined sector have
concentrated on the Born term 
$q\bar{q} \rightarrow e^{+}e^{-}$ to estimate the differential rate  
$\frac{d^4 R_{e^+ e^-}}{d^4 q}$\cite {shu80,kaj86}. In those, the focus has
usually 
been more on the effect of the space time evolution of the plasma 
on the final spectrum. 
Higher order rates have also become recently available \cite{aur02}.
All these rates essentially consist of vacuum processes that 
have been generalized to include medium effects of incoming medium particles along 
with Pauli blocking (Bose enhancement) for outgoing fermions (bosons). 
They also include thermally generated widths and masses for the propagating particles. 
However, these rates have non-zero vacuum counterparts.
Contrary to these are the {\it pure medium} reactions, \ie processes whose vacuum 
counterparts are identically zero. Such processes arise as a result of the 
medium breaking various symmetries which are manifest in the vacuum \cite{chi77wel92}. 
The motivation behind exploring such processes is the possibility of 
observing a spectrum (emanating from these) 
which is noticeably distinct experimentally from the case where a QGP 
was not produced in a collision, or the symmetry remained unbroken by the plasma. 
Such a channel will be explored in this article.

It is now well established that the central region at RHIC is not just heated 
vacuum, but actually displays a finite baryon density \cite{ols01} or an 
asymmetry between baryon and anti-baryon populations. 
This asymmetry may be achieved by the introduction of a quark (or baryon) chemical
potential $\mu_{q}$. For example, 
it may be argued that any baryon number asymmetry prevalent in the 
QGP must have been introduced by valence quarks, which, having encountered 
a hard scattering, failed to exit the central region. Hence, a chemical potential 
$\mu$ is provided for the up and the down quark. The strange quarks are brought 
in by the sea or produced thermally in the medium in equal proportion 
with anti-strange quarks. Hence, they are assigned  $\mu=0$. 
In most heavy-ion collisions, the nuclei of choice are rather
large and display isospin asymmetry, \ie there is an asymmetry in the 
populations of neutrons and protons being brought into the central 
region. If the stranded valence quarks in the plasma arrive with 
equal probability from either nucleon, one would require a higher 
$\mu$ for down quarks. As a first approximation, this effect is ignored, and, 
in the remaining, accept $\mu_u = \mu_d$. 

In such a scenario a finite baryon density may lead to a finite charge density,
this is discussed briefly in Sec. II. It has been proposed that the presence 
of a finite charge density will lead to a new channel for the production of 
lepton pairs \cite{maj01}. Diagrammatically, this is achieved through a
two-gluon-photon vertex with a quark triangle (see Fig. \ref{2vert}). 
The vacuum counterpart of this process is constrained
by Furry's theorem \cite{fur37} and is identically zero. The extension of
this symmetry and its breaking by
the medium was discussed in \cite{maj01b}; for completeness a 
discussion is included in Sec. II. Calculations here will be carried out in the 
imaginary time formalism \cite{kap89}, however our method of treating  finite density 
will differ slightly from the standard method. This is outlined in Sec. III. 
Our formalism leads naturally to the spectator interpretation \cite{won01} of the 
quark loop, also discussed in Sec. III.    
In vacuum, gluon fusion is also constrained by Yang's 
theorem \cite{yan50}. This constraint, based on rotational invariance also 
sets the vacuum counterpart to zero. The extension and 
eventual breaking of this symmetry in the medium are discussed in
Sec. IV. Yang's theorem is broken by two different medium effects: each 
is isolated and the dilepton rate from it is evaluated  in Secs. V and
VI. Concluding discussions are presented in Sec. VII. 
A brief appendix of intermediate derivations follow.


\section{Baryon density, charge density and Furry's theorem}

\begin{figure}
\begin{center}
\hspace{0cm}
  \resizebox{4in}{2in}{\includegraphics[0in,4in][8in,9in]{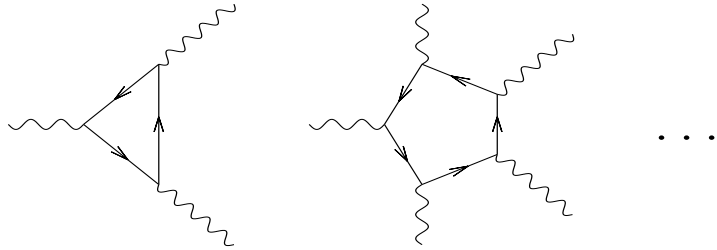}} 
\vspace{0.25cm}
    \caption{ Diagrams that are zero by Furry's theorem and extensions
thereof at finite temperature. These become non-zero at finite charge density.  }
    \label{furry}
  \end{center}
\end{figure}
\begin{figure}
  \begin{center}
\hspace{0cm}
\resizebox{4in}{2.5in}{\includegraphics[0in,4in][8in,9in]{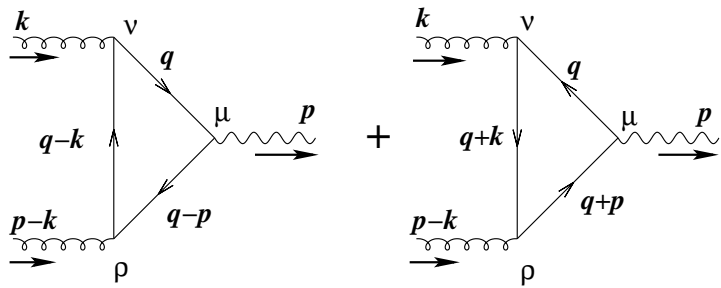}}    
\vspace*{0.3cm}
    \caption{ The two gluon photon effective vertex as the sum of two diagrams
with quark number running in opposite directions. }
    \label{2vert}
  \end{center}
\end{figure}

At zero temperature, and at finite temperature with zero charge density, diagrams 
in QED that
contain a fermion loop with an odd number of photon 
vertices (\emph{e.g.}, Fig. \ref{furry})
are canceled by an equal and opposite contribution coming from the same
diagram with fermion lines running in the opposite direction, this is the 
basic content of Furry's
 theorem \cite{fur37} (see also \cite{itz80,wei95}). 
This statement can also be generalized to QCD for processes 
with two gluons and an odd number of photon vertices. The theorem is 
based solely on charge conjugation invariance of the theory. 

In the language of operators, it may be noted that these 
diagrams are encountered in the perturbative evaluation of Green's 
functions with an odd  number of gauge field operators \ie 

\[
 \langle 0| A_{\mu1} A_{\mu2} ... A_{\mu2n+1}
 |0\rangle . 
\]
 
\nt In QED, $CA_{\mu}C^{-1} = -A_{\mu} $, where $C$ is the 
charge conjugation operator. In the case of the
vacuum, $C|0\rangle = |0\rangle$.  As a result 

\bea
\langle 0| A_{\mu_1} A_{\mu_2} ... A_{\mu_{2n+1}} |0\rangle 
&=&  \langle 0| C^{-1}C A_{\mu_1} C^{-1}C A_{\mu_2} ... 
A_{\mu_{2n+1}} C^{-1} C |0\rangle
\nonumber \\
&=& \langle 0| A_{\mu_1} A_{\mu_2} ... A_{\mu_{2n+1}} |0\rangle (-1)^{2n+1}  \nn \\
&=& -\langle 0| A_{\mu_1} A_{\mu_2} ... A_{\mu_{2n+1}} |0\rangle = 0. \label{fur1}
\eea  

\nt In an equilibrated medium at a temperature $T$, 
we not only have the expectation of 
the operator on the ground state but on all possible matter 
states weighted by a Boltzmann factor \ie 

\[
\sum_{n} \langle n| A_{\mu_1} A_{\mu_2} ... A_{\mu_{2n+1}} |n\rangle 
e^{-\beta (E_n - \mu Q_n)},
\]

\nt where $\beta = 1/T$ and $\mu$ is a chemical potential.
Here, $C|n\rangle = e^{i\phi}|-n\rangle$, where $|-n\rangle$ 
is a state in the ensemble with
the same number of antiparticles as 
there are particles in $|n\rangle$ and vice-versa.
If $\mu = 0$ one obtains


\begin{eqnarray}
\langle n| A_{\mu_1} A_{\mu_2} ... A_{\mu_{2n+1}} |n\rangle 
e^{-\beta E_n}
&=& - \langle -n| A_{\mu_1} A_{\mu_2} ... A_{\mu_{2n+1}} |-n\rangle 
e^{-\beta E_n}. 
\end{eqnarray}

\noindent The sum over all states will contain the mirror term 
$\langle -n| A_{\mu_1} A_{\mu_2} ... A_{\mu_{2n+1}} |-n\rangle e^{-\beta E_n} $, 
with the same thermal weight


\begin{eqnarray}
\Rightarrow \sum_{n} \langle n| A_{\mu_1} A_{\mu_2} ... A_{\mu_{2n+1}} |n\rangle 
e^{-\beta E_n } = 0,
\end{eqnarray}
the expectation over states which are excitations of the vacuum $|0\rangle$ 
will again be zero as in Eq. (\ref{fur1}) and Furry's theorem still holds. 
However, if
$\mu \neq 0$ 


\begin{eqnarray}
\langle n| A_{\mu_1} A_{\mu_2} ... A_{\mu_{2n+1}} |n\rangle 
e^{-\beta (E_n - \mu Q_n)}
= - \langle -n| A_{\mu_1} A_{\mu_2} ... A_{\mu_{2n+1}} |-n\rangle 
e^{-\beta (E_n - \mu Q_n)},
\end{eqnarray}

\noindent the mirror term this time is 
$
 \langle -n| A_{\mu_1} A_{\mu_2} ... A_{\mu_{2n+1}} |-n\rangle 
e^{-\beta (E_n + \mu Q_n)},
$
with a different thermal weight, thus 


\begin{eqnarray}
\sum_{n} \langle n| A_{\mu_1} A_{\mu_2} ... A_{\mu_{2n+1}} |n\rangle 
e^{-\beta (E_n - \mu Q_n)} \neq 0.
\end{eqnarray}
This represents the breaking of Furry's theorem by a medium 
with non-zero charge density or chemical potential.

Some points are in order: there is more than one kind of density that may 
manifest itself in the plasma. There is the net baryon density which requires 
that there be a difference in the populations of quarks and antiquarks of 
a given flavour. There is the net charge density which simply requires that there 
be more of one kind (either positive or negative) 
of charge carrier in the 
medium. Note that it is possible to have a net 
baryon density and yet no charge density 
and vice-versa as Table~\ref{tab1} indicates. 
 As mentioned in the introduction, it will be assumed that 
there is a net baryon density, which manifests itself solely in the up and 
down flavours of the quarks.  As the up quark has a charge of $+\frac{2}{3}$ and the 
down quark $-\frac{1}{3}$; equal densities of both will lead to a QGP
with a net electric charge density. It will be demonstrated in the
following sections that it is this density that leads to a dilepton
signature of the breaking of Furry's theorem at leading order in the EM 
coupling constant. The baryon
density merely serves the purpose of generating such a charge density. 
Hence, this signal is not present, at leading order, in a plasma with a
$\mu_u = \mu_d = \mu_s$, where the net charge is zero
\footnote{An important caveat to this statement is the case where 
each of the quarks has a different mass, 
in which case the effect of one charge carrier may once 
again dominate over another making the signal non-zero.}.

\begin{table}[!htb] 
\begin{tabular}{| l | l | l | l | l | l | r | r |}
\hline
No. of & No. of & No. of & No. of & No. of & No. of & Baryon & Charge \\
$u$'s & $\bar{u}$'s & $d$'s & $\bar{d}$'s & $s$'s & $\bar{s}$'s & density  & density \\
\hline 
\hline
n & n & n & n & n & n & 0        & 0          \\
n & n & 0 & 0 & 0 & 0 & 0        & 0          \\
n & m & 0 & 0 & 0 & 0 & (n-m)/3  & 2(n - m)/3 \\
0 & 0 & n & m & 0 & 0 & (n-m)/3  & (m-n)/3    \\
n & m & n & m & 0 & 0 & 2(n-m)/3 & (n-m)/3    \\
n & m & n & m & n & m &  (n-m)   &  0         \\
n & m & m & n & 0 & 0 &  0       & (n - m )   \\
\hline
\end{tabular}
\caption{ Different scenarios of plasmas with different baryon and charge densities. }
\label{tab1}
\end{table}

In the previous paragraph, pure QED diagrams have been discussed . 
One may now make the most simple extension to QCD, by replacing 
two of the photons with incoming gluons. 
It is to be noted that while the photon is an eigenstate of the charge conjugation 
operator $C$ the gluon is not \cite{pdg}. There are eight gluons, each carrying 
a colour charge in the adjoint representation of SU(3). The sole role 
played by colour in this calculation will be to furnish the factor of Tr$[t^a t^b]$ in 
the Feynman rules. The calculations are identical 
to those in QED. The reasons for considering this sort of diagram over others are 
obvious: this is the lowest order effect in the series, loops with more 
particles attached will invariably be suppressed by coupling constants and 
phase space arguments. Also, 
diagrams with more than two gluons are non-zero in the vacuum itself and 
finite density effects may then be a mere excess on top of an already 
non-zero contribution. Our exploratory calculation mainly seeks to highlight
the behaviour of a new channel.

Cases with all values of the 
three-momentum $\vp$ of the photon from zero (maximum timelike)
up to almost the energy $E$ of the photon (almost lightlike) will be considered. 
We will consider cases where the gluons will be both 
massive and massless. The quarks will be massive in all cases.
\section{Formal calculation}

In this section, the computation of 
the dilepton production rate from the two-gluon 
channel is outlined. The first step is the evaluation of the 
two-gluon-photon vertex in the imaginary time
formalism. This is then used to construct a 
three-loop photon self-energy. The imaginary photon 
frequency is then analytically continued  to 
real values. On the real axis one encounters 
various branch cuts: we pick the cut that 
corresponds to the process of two-gluon fusion and
evaluate the imaginary part of the photon 
self-energy. This is a rather technical procedure. 
However, a method has been proposed 
which allows one to construct the imaginary 
part of the photon self-energy in terms of 
multiple scattering diagrams \cite{won01}. 
This technique has been loosely termed as the 
``spectator interpretation'' of self-energies.
A detailed investigation of this procedure 
has been carried out for two loop 
self-energies in $\phi^4$ theory 
\cite{kap01} as well as in QCD \cite{maj02}. 
In the following subsections the 
spectator interpretation will be extended to three loops. This 
represents a significant application of the spectator 
interpretation, and this permits a 
physical explanation of the extension and 
breaking of Yang's theorem in the medium.  

\subsection{The two-gluon-photon vertex at $\mu \ne 0$ and $\vp \ne 0$ }
 
In the following, we will outline the formal derivation of the 
two-gluon-photon vertex.  Details of the method are 
presented in the appendix.   
The Feynman diagrams for the two-gluon-photon 
vertex as illustrated in Fig. \ref{2vert},
consists of two sets of quark triangle diagrams with the fermion 
number running in opposite directions.
The two vertices are:

\bea 
\mt^{\mu \nu \rho} &=& \frac{-1}{\B} \int \frac{d^3 q}{ (2 \pi)^3 } 
\sum_n \mbox{Tr} \llb i e \kd_{ki} 
\g^\mu  \frac{i ( \f\fq + m ) }{ \fq^2 - m^2 } i g t^b_{ij} \g^\nu \nn \\
&\times& \frac{i ( \f\fq - \f\fk + m ) }{ (\fq-\fk)^2 - m^2 } i g t^c_{jk} \g^\ro 
\frac{ i ( \f\fq - \f\fp + m ) }{ (\fq-\fp)^2 - m^2 }  \lrb \label{vert1} 
\eea

\bea 
\mt^{\mu \ro \nu} &=& \frac{-1}{\B} \int \frac{d^3 q}{ (2 \pi)^3 } 
\sum_n \mbox{Tr} \llb i e \kd_{ik} 
\g^\mu  \frac{i ( \f\fq + \f\fp + m ) }{ (\fq+\fp)^2 - m^2 } i g t^c_{kj} \g^\ro \nn \\
&\times& \frac{i ( \f\fq + \f\fk + m ) }{ (\fq+\fk)^2 - m^2 } i g t^b_{ji} \g^\nu 
\frac{ i ( \f\fq + m ) }{ \fq^2 - m^2 }  \lrb  \label{vert2}\ ,
\eea 
where the trace is implied over both colour and spin indices. 
As always in the imaginary time formalism, 
the zeroth components of each four-momentum is a discrete even or odd 
frequency:
\[
q^0 = i(2n+1)\pi T + \mu , \mbox{\hspace{1cm}} p^0 = i2m \pi T, 
\mbox{\hspace{1cm}} k^0 = i 2j \pi T . 
\]
$n, m, j$ are integers, $\mu$  is the quark chemical potential, and the $t$'s 
are Gell-Mann matrices.  The overall minus sign is 
due to the fermion loop. The sum over $n$ runs over all integers from $-\infty$ to $+\infty$. This 
sum may be performed by two distinct methods: the method of contour integration \cite{kap89} and the 
method of non-covariant propagators \cite{pis88}. Each method is more advantageous in certain 
cases. 
In this article, we use the method of contour integration to evaluate Eqs.
(\ref{vert1}) and (\ref{vert2}) (for an evaluation of similar diagrams using non-covariant propagators 
see Ref. \cite{maj01}).
For later convenience, we will separate the momentum dependent and mass dependent parts of the 
numerators of both  $\mt^{\mu \nu \rho}$ and $\mt^{\mu \rho \nu}$. This is merely a formal 
procedure and for $\mt^{\mu \nu \rho}$ consists of the following:

\bea
\mt^{\mu \nu \rho} &=& \mb^{\mu \A \nu \B \ro \g} {\mathcal{T}_{1}}_{\A \B \g } 
+ \mathcal{A}_1^{\mu \nu \ro}
= \frac{ e g^2 \kd^{bc} }{2 \B} \int \frac{ d^3 q }{ (2 \pi)^3 } \sum_n \mbox{Tr}  \nn \\
& & \Bigg[ \frac{ \mb^{\mu \A \nu \B  \ro \g} q_\A ( q - k )_\B ( q - p )_\g  }
{ ( \fq^2 - m^2 )  ( (\fq - \fk)^2 - m^2 )  ( (\fq - \fp)^2 - m^2 ) } \nn \\
&+& m^2 
\frac{ \ma^{\mu \A \nu \ro} q_\A + \ma^{\mu \nu \B \ro} ( q - k )_\B + \ma^{\mu \nu \ro \g} ( q - p )_\g}
{  ( \fq^2 - m^2 )  ( ( \fq - \fk )^2 - m^2 )  ( ( \fq - \fp )^2 - m^2 )  } \Bigg].
\eea
%
%

\nt Where $\ma^{\mu \nu \ro \g}$ represents the trace of four $\g$ matrices and 
$\mb^{\mu \A \nu \B \ro \g}$ represents the trace of six $\g$ matrices. The denominators of 
both  ${\mathcal{T}_{1}}_{\A \B \g }$ and $\mathcal{A}_1^{\mu \nu \ro}$ are the same and 
hence have the same set of poles.

The sum over 
$n$ may be formally rewritten as a contour integration over an infinite set of contours 
each encircling the points $q^0 = i(2n+1)\pi T + \mu$. The difference between this situation and that 
at zero density ($\mu=0$) is that the contours are on a line displaced by $\mu$ from the $y$ axis. 
In the usual procedure (see Sec. (3.6) of Ref. \cite{kap89}), one separates the vacuum piece, a thermal 
particle and antiparticle piece, and a pure density contribution. However it is possible to deform the 
contours in a way entirely similar to the zero density situation. One obtains two 
infinite closed semi-circular contours: see appendix for details. 
The sole difference from the zero density situation is that
one of the contours will be multiply connected: in the case of $\mt^{\mu \nu \rho}$ this consists of the 
exclusion of the points at $q^0 = i(2n+1)\pi T + \mu$ by an infinite set of infinitesimal contours, 
while in the case of  $\mt^{\mu \rho \nu}$, the points at $q^0 = i(2n+1)\pi T - \mu$ 
are excluded. Performing the contour integration will essentially result in the 
evaluation of the 
residues of the functions in Eqs. (\ref{vert1}) and (\ref{vert2}) at its various poles with appropriate 
finite density thermal weights. Combining the results obtained from the application of this procedure 
on $\mt^{\mu \nu \rho}$ and $\mt^{\mu \rho \nu}$ one obtains

\bea
T^{\mu \nu \rho} &=& \int \frac{ d^3 q }{ (2 \pi)^3 } \sum_{i} \llb \h(\w_i) 
\left( \frac{1}{ e^{ \B ( q^0 - \mu )  }  +  1 } - \frac{1}{ e^{ \B ( q^0 + \mu )  }  +  1 } \right) 
+ \h(-\w_i) \left( \frac{1}{ e^{ \B ( -q^0 - \mu )  }  +  1 } - 
\frac{1}{ e^{ \B ( -q^0 + \mu )  }  +  1 } \right)
\lrb  \nn \\
&\times& \frac{ e g^2 \kd^{bc} }{2 \B}   \res   
\Bigg[ \frac{ \mb^{\mu \nu \ro}_{\A \B \g } q^\A ( q - k )^\B ( q - p )^\g  }
{ ( \fq^2 - m^2 )  ( (\fq - \fk)^2 - m^2 )  ( (\fq - \fp)^2 - m^2 ) } \nn \\
&+& 4m^2 \frac{ \gmn ( q - p - k )^\ro  + \gmr ( q - k + p )^\nu  +  \gnr ( q + k - p )^\mu }
{  ( \fq^2 - m^2 )  ( ( \fq - \fk )^2 - m^2 )  ( ( \fq - \fp )^2 - m^2 )  } \Bigg] \Bigg|_{q^0=\w_i}
\label{vert3,1}
\eea

\nt 
Where, the $w_i$'s (with $i$ running from 1 to 6) 
are the residues of the function within the large square brackets. 
We find, as would have been expected, that the entire contribution is proportional to the 
difference of the quark and antiquark distribution functions. We denote these as
$\kD \nf(q^0 , \mu) =  
( \frac{1}{ e^{ \B ( q^0 - \mu )  }  +  1 } - \frac{1}{ e^{ \B ( q^0 + \mu )  }  +  1 })$. 
The residues will be evaluated at the various poles of the integrand.
A close inspection of Eq. (\ref{vert3,1}) indicates that there are three poles on the 
positive $x$ axis at, 

\bea
q^0 &=& \sqrt{q^2 + m^2} = E_q \label{res1}\\
q^0 &=& \sqrt{ | \vq - \vk |^2 + m^2 } + k^0 = E_{q-k} + k^0 \\
q^0 &=& \sqrt{ | \vq - \vp |^2 + m^2 } + p^0 = E_{q-p} + p^0. 
\eea
 
\nt and three on the negative $x$ axis,

\bea
q^0 &=& -\sqrt{q^2 + m^2} = -E_q \\
q^0 &=& -\sqrt{ | \vq - \vk |^2 + m^2 } + k^0 = -E_{q-k} + k^0 \\
q^0 &=& -\sqrt{ | \vq - \vp |^2 + m^2 } + p^0 = -E_{q-p} + p^0. \label{res2}
\eea
We denote the residue at each of these poles as residues (1-6). Before evaluating 
the function at each of these residues, we consider the fate of the remaining 
imaginary frequencies in the expressions $k^0,p^0$. The even frequency $k^0$ 
also has to be summed in similar fashion as $q^0$. The external photon frequency
$p^0$ will have to be analytically continued to a general complex value and finally the 
discontinuity of the full self-energy across the real axis of $p^0$ will be considered. We 
perform this procedure in the next section.  

\subsection{The photon self-energy and its imaginary part}

\begin{figure}[!htb]
  \begin{center}
  \epsfxsize 70mm
\hspace{1.5cm}  
\epsfbox{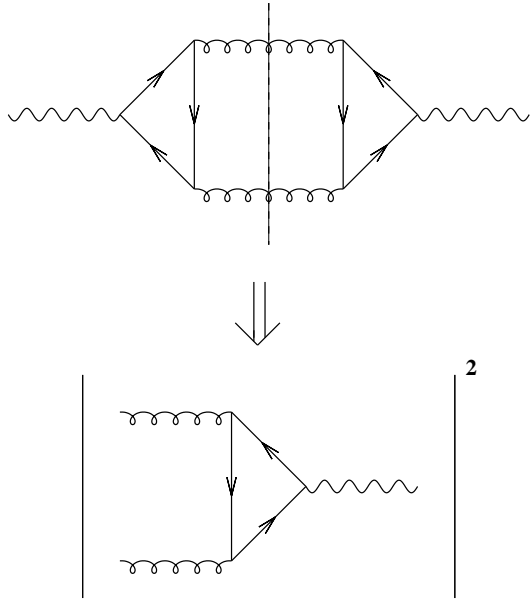}
    \caption{ The Full photon self-energy at three loops and the cut that is
evaluated. }
    \label{dselfE}
  \end{center}
\end{figure}

We are now in a position to calculate the contribution made by the diagram of
Fig. \ref{2vert}
to the dilepton spectrum emanating from a quark gluon plasma. 
To achieve this, we choose
to calculate the discontinuity  of the photon self-energy as represented by the diagram of
Fig. \ref{dselfE} across the real axis of $p^0$. 
In the previous section we wrote down expressions for 
$T^{\mu \nu \ro}(p,k,p-k)$: the vertex with the the two gluon momenta incoming and 
the photon momentum outgoing. To write down the expression for the full 
self-energy we also need expressions for ${T'}^{\mu \nu \ro}(-p,-k,k-p)$: the 
vertex with the photon momentum incoming and the gluon momenta outgoing. This 
vertex also admits a decomposition into two pieces for quark number running in 
opposite directions,

\begin{equation}
{T'}^{\mu \nu \ro}(-p,-k,k-p) = {\mt'}^{\mu \nu \ro} + {\mt'}^{\mu \ro \nu} ,  \label{t1'=a+b}
\end{equation}

where the factor ${\mt'}^{\mu \nu \ro}$ can be written as 

\bea
{\mt'}^{\mu \nu \rho} &=& \mb^{\mu \g \nu \B \ro \A } \mathcal{T'_1}_{\g \B \A } 
+ \mathcal{A'}_1^{\mu \nu \ro}
= \frac{ e g^2 \kd^{bc} }{2 \B} \int \frac{ d^3 q }{ (2 \pi)^3 } \sum_n   \nn \\
& & \Bigg[ \frac{ \mb^{\mu \g \nu \B \ro \A} q_\g ( q + k )_\B ( q + p )_\A  }
{ ( \fq^2 - m^2 )  ( (\fq + \fk)^2 - m^2 )  ( (\fq + \fp)^2 - m^2 ) } \nn \\
&+& \frac{ \ma^{\mu \g \nu \ro} q_\g + \ma^{\mu \nu \B \ro} ( q + k )_\B + \ma^{\mu \nu \ro \A} ( q + p )_\A }
{  ( \fq^2 - m^2 )  ( ( \fq + \fk )^2 - m^2 )  ( ( \fq + \fp )^2 - m^2 )  } \Bigg] \label{vert1'}
\eea

\nt The traces of four and six $\g$ matrices admit the following identities:

\bea 
\ma^{\mu \g \nu \ro} &=& \ma^{\ro \nu \g \mu} = \ma^{\mu \ro \nu \g} \\
\ma^{\mu \nu \B \ro} &=& \ma^{\ro \B \nu \mu} = \ma^{\mu \ro \B \nu} \\
\ma^{\mu \nu \ro \A} &=& \ma^{\A \ro \nu \mu} = \ma^{\mu \A \ro \nu} \\
\mb^{\mu \g \nu \B \ro \A} &=& \mb^{\A \ro \B \nu \g \mu} = \mb^{\mu \A \ro \B \nu \g}
\eea
In each equation above the first equality uses the fact that the trace of 
$n$ $\g$ matrices in 
a particular order is the same if the order is fully reversed.
The second equality uses the cyclic properties of the trace to put $\g^\mu$ 
at the start in each case. Substituting the above identities in 
Eqs.~(\ref{vert1'}) and (\ref{t1'=a+b}), we may easily demonstrate that

\begin{equation}
{T'}^{\mu \nu \ro}(-p,-k,k-p) = T^{\mu \nu \ro}(p,k,p-k)
\end{equation}

Implementing the above simplifications we may, formally, write down the 
full expression for the three-loop photon self-energy in the imaginary time formalism as 
(note, there are many other photon self-energy diagrams at three loops, 
however we only consider the contribution which arises due to the breaking of 
$C$ invariance) 

\bea
i\Pi^{\mu \nu}(p) &=& \frac{i}{\beta} \sum_{k^{0}} \int \frac{d^{3}k}{(2\pi)^{3}} 
{iT'}^{\mu \ro \g}(-p,-k,k-p) {\mathcal D}_{\ro \zeta}(k)  
iT^{\nu \zeta \kd}(p,k,p-k) {\mathcal D}_{\kd \g}(p-k) \nn \\
&=& \frac{i}{\beta} \sum_{k^{0}} \int \frac{d^{3}k}{(2\pi)^{3}} 
iT^{\mu \ro \g}(p,k,p-k) {\mathcal D}_{\ro \zeta}(k)  
iT^{\nu \zeta \kd}(p,k,p-k) {\mathcal D}_{\kd \g}(p-k) .  \label{fullselfE}
\eea

\nt The diagram that we are considering is that of Fig. \ref{dselfE}. 
In the above equation ${\mathcal D}_{\ro \zeta}(k)$ is the gluon propagator. 
We perform the calculation in the Feynman gauge for the gluons:

\begin{equation}
{\mathcal D}_{\ro \zeta}(k) = \frac{-i g_{\ro \zeta}}{k^{2}}.
\end{equation}

\noindent 
In order to calculate the differential rate of dilepton production we need to evaluate
the discontinuity of the photon self-energy. This involves, first, 
converting the sum over discrete $k^0$ frequencies into a contour
integral over a complex continuous $k^0$, as was done for $q^0$. This would be 
followed by the evaluation of the contour integral by summing over the residues of the integrand at
each of the poles of $k^0$. Finally we 
look for poles and branch cuts in this expression as a function of $p^{0}$ by
analytically continuing $p^{0}$ onto the real axis. There are many poles in $k^0$ 
for which we have to evaluate residues. Some of these poles are in the 
denominators of the gluon propagators, 
while some are in the vertices $T^{\mu \nu \ro}$ and ${T'}^{\mu \nu \rho}$. 
As the residues at each of these poles is analytically 
continued in $p^0$ from a discrete imaginary 
frequency to a complex number and finally to a 
real continuous energy, various branch cuts will appear.
This procedure of evaluation of residues and analytic continuation may also be 
performed prior to the $d^3 k$ or $d^3 q$
integrations: at this stage the branch cuts on the real $p^0$ axis appear as 
poles in the $d^3 k$ or $d^3 q$ integrations as $p^0 \ra E+i\e$. 
The presence of the $i\e$ will allow each integrand to be unambiguously 
broken up into a set of principle values and imaginary parts. Combining 
these will lead to various real and imaginary parts of the full self-energy and will 
correspond to various physical processes of photon propagation and decay in the medium. 
Twice the integral over the imaginary parts will give us the required discontinuity. 

To obtain the 
discontinuity we essentially choose a pair of poles in the expression; evaluate the 
residue of the $d k^0$ integration at the first pole and twice the imaginary part of  
the $d^3k, d^3q$ integration as $p^0 \ra E+ i \e$ at the second pole. 
Each such combination constitutes a `cut' 
of the self energy or a part of a cut of the self-energy. 
The cut line essentially passes through a set of propagators in the self-energy dividing it
into two disjoint pieces (see, for example, Fig. \ref{dselfE}). The propagators that 
have been cut are indicated by the energy momentum delta functions obtained from the 
residue and discontinuity procedure.
If we denote the Feynman rule for one of the disjoint pieces as $\mat_1$ and the other
by $\mat_2$ then this particular discontinuity of the self-energy gives the Feynman rule for 
$\mat_2^*\mat_1$ or $\mat_1^*\mat_2$. 
If the cut is symmetric \ie $\mat_1 = \mat_2$, then we obtain the square of the 
amplitude for the process $|\mat_1|^2$. For this calculation we are solely interested in the 
square of the amplitude of the process shown in the lower panel of Fig. \ref{dselfE}. 
Our preceding discussion indicates that this will be given by the cut line indicated in the 
upper panel of the figure. 
This is the process of gluon gluon fusion to produce a heavy photon resulting 
in a dilepton. The 
other cuts represent extra finite density contributions to processes already non-vanishing at 
zero density. 

The above discussion indicates that we merely have to look for poles in the denominators of 
the gluon propagators. Isolating this piece form Eq. (\ref{fullselfE}), we note the denominators of 
the two sets of gluon propagators is 

\begin{equation}
\frac{1}{\fk^2} \frac{1}{(\fp - \fk)^2} = \frac{1}{( k^0 - k )( k^0 + k ) } 
\frac{1}{ ( p^0 - k^0 - E_{p-k} )( p^0 - k^0 + E_{p-k} ) }
\end{equation}

\nt Where $E_{p-k} = | \vp - \vk | $. 
The $k^0$ integration will encounter four possible poles at $k^0 = \pm k$, and $k^0 = p^0 \pm E_{p-k}$.
Each choice will lead to a different process as $p^0$ is analytically continued to the real axis. 
All choices 
will not lead to the desired process. 
We now investigate each of these possibilities in turn. 

We begin by evaluating the residue of the remaining integrand at the pole $ k^0 = k $. At this pole the 
remaining three denominators are 

\[
\frac{1}{2k}  \frac{1}{p^0 - k - \pk }   \frac{1}{p^0 - k + \pk }.
\] 

\nt On analytically continuing $p^0$ we will obtain two locations on the real line of $p^0$  
where a discontinuity may occur: 
$p^0 = E = k + \pk $ and $p^0 = k - \pk $. The second pole will lead to the 
photon invariant mass $E^2-p^2 < 0$  \ie a spacelike photon, we ignore this cut.
Substituting the first value for $p^0$, we obtain the discontinuity of the 
self energy at $E = k + \pk$. This turns the gluon denominators into 

\[
-i \pi \kd(E-k-\pk) \frac{1}{2k} \frac{1}{2\pk} .
\]

Evaluating the residue at $k^0 = -k $, we obtain the remaining denominators as
\[
\frac{1}{-2k} \frac{1}{-p^0 - k - \pk } \frac{1}{-p^0 - k + \pk }.
\] 
This leads to two possible locations on the real line of $p^0$ 
where a discontinuity may occur: 
$ -k - \pk$ and $\pk - k$. The first choice leads to a
negative energy photon and the second to a photon with a spacelike invariant mass, 
thus we ignore this 
$k^0$ pole altogether.

Evaluating the residue at $k^0 = p^0 + \pk$ and analytically continuing $p^0$, 
we once again obtain a negative energy photon and a
spacelike photon and thus this residue is ignored as well. The final residue is 
at $k^0 = p^0 - \pk$. This leads to possible discontinuities at $p^0 = E = k + \pk$
and $\pk - k$. The second possibility leads to a spacelike photon and is ignored.
The first gives a timelike photon with positive energy and thus is included in the 
cuts considered. With this choice we obtain the gluon denominators into 
\[
-i \pi \kd(E-k - \pk) \frac{1}{2k} \frac{1}{-2\pk}
\] 

Thus, in performing the sum over the Matsubara frequencies $k^0$ we will only confine ourselves to
two poles: one on the positive side of the real axis at $k^0 = k$, one on the negative 
side at $k^0= p^0 - \pk$. 
For both poles, we analytically continue $p^0$ to $E = k + \pk$, leading to 
\[
k^0 = p^0 - \pk = E - \pk = k + \pk - \pk = k. 
\]
Thus, in the rest of the expression we will simply replace $k^0 \ra k$ and use the appropriate 
distribution functions in each case depending on whether the initial $k^0$ pole was on the positive or 
negative side. Then we will use the delta function to set the value of $k$. The results of this 
procedure as well as the final expressions and their properties will be discussed in the next subsection. 

One may also expect the gluons to acquire a thermal 
dispersion relation in the hot QCD medium (see Ref. \cite{leb96}). 
In a later section we will employ a simplified version of the in-medium 
gluon dispersion relations: the gluons will be ascribed a thermal mass. 
The above derivation of the photon self-energy and the pole 
analysis are still valid provided we use a massive vector propagator such as:
\begin{equation}
{\mathcal D}_{\ro \zeta}(k) = -i\frac{ g_{\ro \zeta}-\frac{k^{\ro}k^{\zeta}}{m_g^2}}{k^{2}-m_g^2}.
\end{equation}
and we substitute in the vertex expressions every occurrence 
of the massless gluon energy $k$ by its massive equivalent 
$E_k =\sqrt{k^2+m_g^2}$ where $m_g$ is the thermal gluon mass.

\subsection {The spectator interpretation}    

In this section we evaluate the particular cut of Fig. \ref{dselfE} of the three-loop 
photon self-energy. Focusing on the two poles of $k^0$ highlighted in 
the preceding subsection and performing the associated analytic continuation
of $p^0$ we obtain the discontinuity in the photon self-energy as

\bea 
\dsc \Pi^{\mu \nu} &=&  \int \frac{d^{3}k}{(2\pi)^{3}} 
T^{\mu \ro \g}(E,k,\pk) \frac{ g_{\ro \zeta} }{2k}  
T^{\nu \zeta \kd}(E,k,\pk) \frac{g_{\kd \g}}{2\pk} \nn \\
& & \times \left[ \af + \frac{1}{e^{ \B k } - 1 } \right] 
(-1) \Big( -2 \pi i \kd(E - k - \pk) \Big) \nn \\ 
&-& \int \frac{d^{3}k}{(2\pi)^{3}} 
T^{\mu \ro \g}(E,k,\pk) \frac{ g_{\ro \zeta} }{2k}  
T^{\nu \zeta \kd}(E,k,\pk) \frac{g_{\kd \g}}{2\pk} \nn \\
& & \times \left[ \af + \frac{1}{e^{ \B (-\pk) } - 1 } \right] 
(-1) \Big( -2 \pi i \kd(E - k - \pk) \Big) \nn \\ 
\eea

\nt Combining the gluon distribution functions and using the relation $\pk + k = E$, we obtain 

\bea 
\dsc \Pi^{\mu \nu} &=&  \int \frac{d^{3}k}{(2\pi)^{3}} 
T^{\mu \ro \g}(E,k,\pk) \frac{ g_{\ro \zeta} }{2k}  
T^{\nu \zeta \kd}(E,k,\pk) \frac{g_{\kd \g}}{2\pk}  \nn \\
& & \times \bigg( e^{\B E} - 1 \bigg) n(k) n(\pk)  \Big( 2 \pi i \kd(E - k - \pk) \Big) 
\eea

To obtain the differential rate for dilepton production we need the quantity 
$ r = \left[ \frac{p_\mu p_\nu}{\fp^2} - g_{\mu \nu} \right] \dsc[-i \Pi^{\mu \nu}]$. We substitute the 
expression for $\dsc [\Pi^{\mu \nu}]$  and note that the intervening factors of the metric as well 
the factor $ \frac{p_\mu p_\nu}{\fp^2} - g_{\mu \nu}$ are obtained from the sum over the 
polarizations of the incoming gluons and the outgoing photon:

\begin{equation}
\sum_{i} {\ve_i^*}_\ro (\fk) {\ve_i}_\zeta (\fk) \ra - g_{\ro \zeta}
\end{equation}

\begin{equation}
\sum_{l} {\ve_l}_\mu (\fp) {\ve^*_l}_\nu (\fp) =   \frac{p_\mu p_\nu}{\fp^2} - g_{\mu \nu}
\end{equation}

\nt Substituting the above relations into $r$, we obtain

\bea 
r &=& \sum_{i} \sum_{j} \sum_{l}  \int \frac{d^{3}k}{(2\pi)^{3} 2k 2\pk} 
\Big[ {\ve_l}_\mu T^{\mu \ro \g} {\ve_i^*}_\ro {\ve^*_j}_\g \Big]
\Big[ {\ve^*_l}_\nu T^{\nu \zeta \kd} {\ve_j}_\kd {\ve_i}_\zeta \Big] \nn \\
& & \times \bigg( e^{\B E} - 1 \bigg) n(k) n(\pk)  \Big( 2 \pi \kd(E - k - \pk) \Big) \label{r}
\eea

\nt Introducing factors of $2 \pi$ and extra delta functions we may formally write the 
 above as a straightforward kinetic theory equation: 

\bea 
r &=& \sum_{i} \sum_{j} \sum_{l}  \int 
\frac{d^{3}k}{(2\pi)^{3} 2k}  \int \frac{d^{3} \w}{(2\pi)^{3} 2\w} 
\Big[ \mat_{i,j,l} \Big]^*
\Big[ \mat_{i,j,l} \Big] \nn \\
& & \times \bigg( e^{\B E} - 1 \bigg) n(k) n(\pk)  
\Big( (2 \pi)^4 \kd^4(\fp - \fk - \fw) \Big) \ ,
\label{kin_th}
\eea
where $\mat_{i,j,l}$ is the thermal matrix element for two gluons in polarization states 
$i,j$ to make a transition into a photon in a polarization state $l$. The entire 
process is weighted by the appropriate thermal gluon distribution functions and 
has the usual energy momentum conserving delta function. 
In this calculation both gluons are massless; thus they have only two 
physical polarizations. The photon being massive has an extra polarization $\ve_3^{\mu}$. 
Note that the thermal matrix elements $\mat_{i,j,l}$  still contain thermal distribution
functions, they thus encode information regarding incoming  and outgoing 
quarks from the process into the medium. 
Using the polarization vectors, the thermal matrix elements may be easily expressed as 
vacuum multiple scattering diagrams with thermal weights for the incoming and outgoing 
quarks as well. This reformulation of the thermal loops is the spectator 
interpretation. We now no longer have a quark loop: it   is replaced with a set of 
coherent tree diagrams.  

\begin{figure}[htb]
  \begin{center}
  \epsfxsize 90mm
\hspace{1.5cm}  
\epsfbox{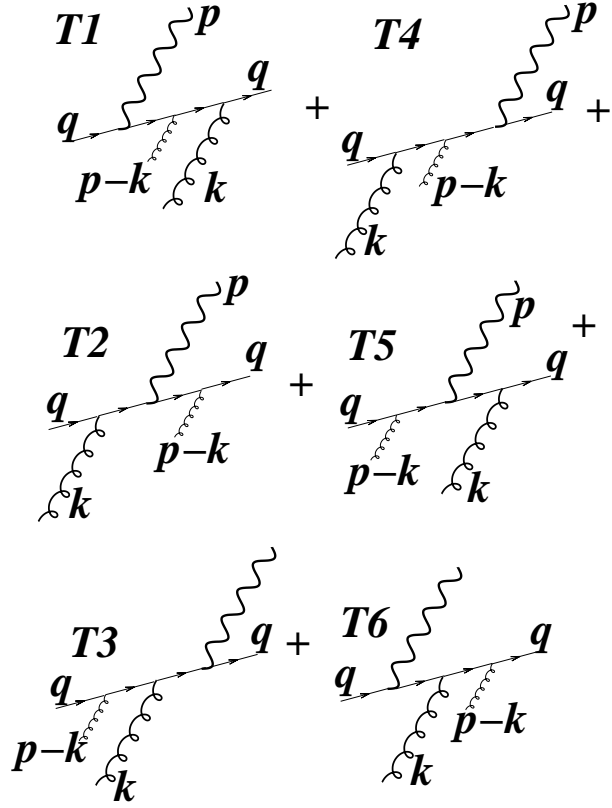}
    \caption{ Multiple scattering expansion of the quark loop. Each diagram
    corresponds to a residue of the $q^0$ integration. }
    \label{spec}
  \end{center}
\end{figure}

Recall that in the 
evaluation of $T^{\mu \nu \ro}$ we had performed a contour integration over
$q^0$  and obtained six residues (see Eqs. (\ref{res1})--(\ref{res2})). 
We did not elucidate the residues at the time, as
we still had two complex frequencies,
$k^0$ and $p^0$, in the expressions. Once the residue in the $k^0$ integration
has been taken and 
$p^0$ analytically continued, we obtain the following results.

For the pole at $q^0= \sqrt{q^2 + m^2} = E_q $ we obtain, 

\bea
T_1^{\mu \nu \rho} (q^0 = E_q, k^0 = k, p^0 = E) &=& \frac{ e g^2 \kd^{bc} }{2}
\int \frac{ d^3 q }{ (2 \pi)^3 } \nn \\ 
&\times& \llb \frac{ \mb^{\mu \nu \ro}_{\A \B \g } q^\A ( q - k )^\B ( q - p )^\g  }
{ 2E_q ( (\fq - \fk)^2 - m^2 )  ( (\fq - \fp)^2 - m^2 ) } \nn \\
&+&  4 m^2  \frac{ \gmn ( q - p - k )^\ro  + \gmr ( q - k + p )^\nu  +  \gnr ( q + k - p )^\mu }
{  2E_q  ( ( \fq - \fk )^2 - m^2 )  ( ( \fq - \fp )^2 - m^2 )  }
\lrb \Delta \nf(E_q,\mu) 
\label{spectator1}
\eea

\nt Once again $ \mb^{\mu \nu \ro}_{\A \B \g }$ represents the trace of 
six gamma matrices. The term above may be reinterpreted as

\bea
T_1^{\mu \nu \rho} &=& \frac{- e g^2 \kd^{bc} }{2}
\int \frac{ d^3 q }{ (2 \pi)^3 2E_q }  
 \sum_{r}  \llb \frac{ \bar{u}_r  (q) \g^{\nu} (\f\fq - \f\fk + m)
\g^{\rho} (\f\fq - \f\fp + m ) \g^{\mu} u_r (q)    }
{ ( (\fq - \fk)^2 - m^2 )  ( (\fq - \fp)^2 - m^2 ) }    \nn \\ 
&\times& \af \llb \Big[ 1 - \nf(E_q,\mu) - \nf (E_q,\mu) \Big]  
- \Big[ 1 - \nf(E_q,-\mu) - \nf (E_q,-\mu) \Big]  \lrb \nn \\
&=& t_1^{\mu \nu \rho} \af \llb \Big[ 1 - \nf(E_q,\mu) - \nf (E_q,\mu) \Big]  
- \Big[ 1 - \nf(E_q,-\mu) - \nf (E_q,-\mu) \Big]  \lrb
\label{spectator1b}
\eea

\nt Where $r$ is the spin of the quark (or antiquark) of momentum $q$. The
distribution functions have been written in a way to distinguish the
contributions from quarks and antiquarks. If we concentrate only on the 
coefficient of the quark part of the distribution functions we note that 
$\mps^1_{i,j,l} = {\ve_l}_\mu t_1^{\mu \ro \g} {\ve_i^*}_\ro {\ve^*_j}_\g $
is simply the Feynman rule for the process indicated as $T1$ in
Fig. \ref{spec}. The spins for the incoming quark have been averaged over, 
while its momentum has been integrated over. One may also show that the 
coefficient of the antiquark part of the distribution functions corresponds 
to the diagram referred to as $T4$ in Fig. \ref{spec}; with the in
coming quark line replaced by an incoming antiquark line. 

Following the procedure, as outlined above, one may easily demonstrate that 
each residue of $q^0$ corresponds to a multiple scattering 
topology. As a result there are six different 
coherent tree diagrams as shown in Fig. \ref{spec}. 
Each tree diagram in Fig. \ref{spec} corresponds to a residue of the $q^0$
integration. No particular time ordering is implied except that the gluons are
incoming and the photon is outgoing. The quarks can be both incoming and out
going. As a result the thermal matrix element $\mat_{i,j,l}$ in
Eq. (\ref{kin_th}) may be expanded as

\bea
\mat_{i,j,l} &=& \af \left( \mps_{i,j,l}^1 + \mps_{i,j,l}^2 + \mps_{i,j,l}^3 + \mps_{i,j,l}^4 +
\mps_{i,j,l}^5  + \mps_{i,j,l}^6 \right) \Bigg[ \left\{1 - \nf(E_q,\mu) \right\} -
\nf(E_q,\mu)  \Bigg] \nn \\
& & + \af \left( \mas_{i,j,l}^1 + \mas_{i,j,l}^2 + \mas_{i,j,l}^3 + \mas_{i,j,l}^4 +
\mas_{i,j,l}^5  + \mas_{i,j,l}^6 \right) \Bigg[ \left\{1 - \nf(E_q,-\mu) \right\} -
\nf(E_q,-\mu)  \Bigg] \label{mtrx_elm}  \label{spec_eqn}
\eea  

\nt Where $\mps_{i,j,l}^1$ is the vacuum amplitude of the diagram referred to as
$T1$ in Fig. \ref{spec}, and $\mas_{i,j,l}^1$ is the same diagram with the 
in and outgoing quark replaced with an antiquark. It may be easily demonstrated, by a 
simple variable transformation, that 
$\mas_{i,j,l}^1 = -\mps_{i,j,l}^4$. The same is true for the other amplitudes 
with antiquarks, each is the negative of a different vacuum amplitude from the 
six. As a result we obtain:

\bea
\mat_{i,j,l} &=& \left( \mps_{i,j,l}^1 + \mps_{i,j,l}^2 + \mps_{i,j,l}^3 + \mps_{i,j,l}^4 +
\mps_{i,j,l}^5  + \mps_{i,j,l}^6 \right) 
\Bigg[ - \nf(E_q,\mu) + \nf(E_q,-\mu)  \Bigg]  \label{spec_eqn_b}
\eea  

\begin{figure}[htb]
  \vspace{-4cm}
 \begin{center}
  \epsfxsize 150mm
\epsfbox{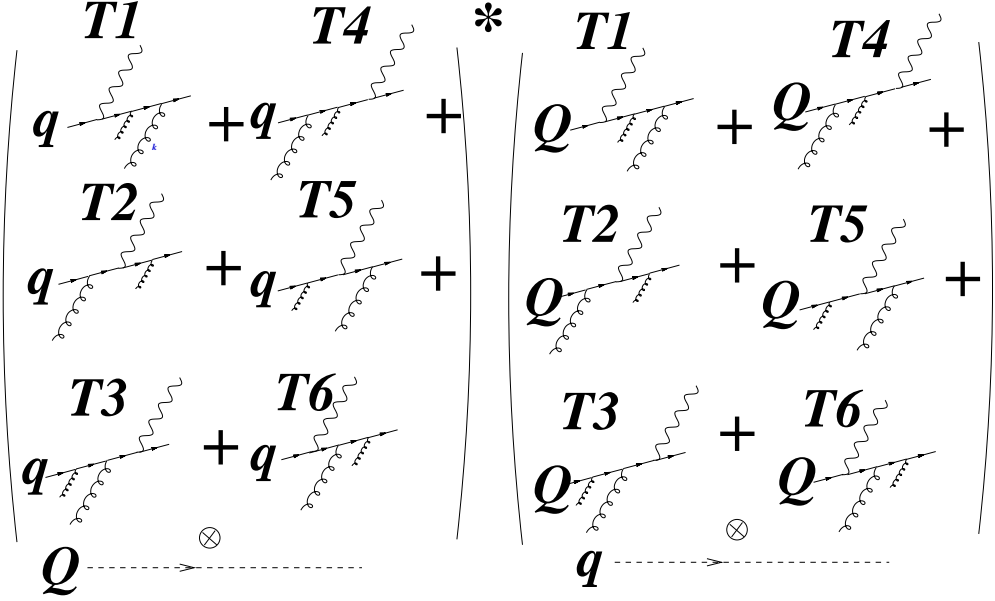}
 \vspace{-4cm}
   \caption{ Multiple scattering expansion of 
   the lower diagram in Fig. \ref{dselfE}. Each diagram
    corresponds to a residue of the $q^0$ integration. }
    \label{spec2}
  \end{center}
\end{figure}

From the above expression it is obvious that if the chemical potential $\mu=0$
then the rate is zero as well. 
Note that the imaginary time formalism only provides us with
the square of the above term. Indeed it is $\mat^2$ which will eventually
determine the rate. The uncoupling of the $\mat^2$ to individual
tree amplitudes constitutes the proof of the spectator interpretation of the 
imaginary time formalism. One may derive the rate starting from these 
simple tree amplitudes, without invoking the complicated machinery of the 
imaginary time formalism. 

We may now state the spectator interpretation for the square of
matrix element. This is shown in Fig. \ref{spec2}. This is the spectator
interpretation of the loop diagram of Fig. \ref{dselfE}. It represents the
process of two gluons in states $(i,j)$  encountering two incoming medium quarks (or
antiquarks) with quantum numbers $q,Q$ leading to the emission of a photon in
state $l$ and two quarks (or antiquarks) with identical quantum numbers $q,Q$. 
In the amplitude on the left hand side of Fig. \ref{spec2} $q$
participates in the reaction whereas $Q$ is a spectator. In the amplitude on
the right the reverse is true.  Note that we do not require $q,Q$ to be
simultaneously quarks or antiquarks, they may be either. We have 
thus expressed the complicated loop containing matrix element as a 
coherent sum of simpler tree diagrams. The main purpose of such a 
decomposition is more than just a physical perspective: it allows us an 
understanding of the mechanism of symmetry breaking not provided by the 
rules of the imaginary time formalism. This will be discussed in the subsequent 
section. 

In passing, we should once again point out how the spectator interpretation 
greatly simplifies any thermal calculation. The diagrams of Fig. \ref{spec} are 
not difficult to motivate from first principles. They represent the set of all possible 
means (at lowest order in coupling) 
by which one may couple two gluons to a photon with two fermions from the 
medium. Along with this is the restriction 
that the fermions return back to the states that they vacated.
The six diagrams represents the six different means of ordering the gluons and the photon. 
The two kinds of thermal factors $(\nf(E_q,\pm\mu) , \left\{1 - \nf(E_q,\pm\mu) \right\})$ 
represent the possibilities of the fermion being ejected 
from the medium prior to its re-absorption, and vice-versa. The two fermions are allowed to 
be both quarks and antiquarks as required by a relativistic medium.

\section{Rotational invariance and Yang's theorem.} 

The vacuum analogue of the two-gluon-virtual photon process does not exist 
due to Furry's theorem. If it did, it would represent 
an instance of two identical massless vectors fusing to 
form a massive spin one object; or alternatively a 
massive spin one object decaying into two massless vectors. 
There exist other such processes not protected by Furry's 
theorem, e.g., $\omega \ra \g \g $. Such a process 
though not blocked by Furry's theorem is still vanishing in 
the vacuum.
We effectively have a situation 
where there are two massless spin one particles in the in state
and a spin one particle in the out state, or vice versa. In such circumstances 
another symmetry principle is invoked. This symmetry principle, 
due to C. N. Yang \cite{yan50}, is based on the parity and rotational symmetries 
of the in and out states and will, henceforth, be referred to as Yang's 
theorem. 


\subsection{Yang's theorem in vacuum}


The basic statement of Yang's theorem, as far as it relates to this 
calculation, is that it is impossible for a spin one particle in 
vacuum to decay into two massless vector particles. This statement 
is obviously also true for the reverse process of two massless 
vectors fusing to produce a spin one object and as a result a 
fermion and an antifermion combination in the triplet state. 
This may be understood 
through the following simple observation. Imagine that we boost to 
the frame where the two incoming vectors (in this case gluons) are 
exactly back-to-back with their three-momenta equal and opposite. The 
outgoing vector (the virtual photon in this case) is produced at rest and 
eventually disintegrates into a lepton pair. We will 
now apply various symmetry operators (parity, rotation, etc.) on both 
the incoming and outgoing states. Note that, as we are only interested in 
strong and electromagnetic interaction, hence, parity is a good quantum 
number. If both incoming and outgoing states 
are found to be eigenstates of the symmetry operator then 
they must be eigenstates with exactly the same eigenvalues, else this 
transition is not allowed.  

We begin the discussion with the parity operator $\mathcal{P}$. We align the 
$z$ axis along the direction of one of the incoming gluons. The outgoing 
or final state is parity-odd, as we know that
our final state is the photon, or a state composed of a 
lepton and anti-lepton in the $^3S$ state.
The gluons, on-shell in this calculation, are each parity-odd. 
We may still construct a parity-odd in state via the following method: we label 
the possible in states as 

\[
| R+ ; R- \rc , \mbox{\hspace{0.25cm}} | L+ ; L- \rc , \mbox{\hspace{0.25cm}} 
| L+ ; R- \rc , 
\mbox{\hspace{0.25cm}}
 | R+ ; L- \rc . 
\]

\nt Where, the $| R+ ; R- \rc $ is the state where both gluons are right handed. The 
$| L+ ; R- \rc $ state indicates that the gluon moving in the positive $z$ direction is
left-handed while that moving in the negative 
$z$ direction is right handed (we have used 
the notation that the $+$ sign indicates the gluon moving in the 
positive $z$ direction). The parity operation interchanges the momenta of the 
two gluons but leaves the direction of their spins intact. Hence the state 
$| R+ R- \rc  - | L+ L- \rc $  is odd under parity operation. This implies that 
only this combination of incoming gluons is allowed by parity to fuse to form 
the virtual photon and hence the lepton pair.

We next turn to the rotation operator, $\mathcal{R}$.
The in state is the state of two gluons; the out state may be 
considered to be either the temporary virtual photon, or the finally 
produced pair of lepton anti-lepton. One may chose 
either for this analysis; we decide on the 
photon as it is simpler. 
For the in state we use the only state that is 
allowed by parity \ie $| R+ R- \rc - | L+ L- \rc $. This 
state may be re-expressed as the action of creation and annihilation 
operators on the vacuum state as,

\bea
| R+ R- \rc - | L+ L- \rc = \left[  \ad_{R+} \ad_{R-}  -  \ad_{L+} \ad_{L-}  \right]  | 0 \rc.
\eea
  
\nt Where $| 0 \rc$ is the vacuum state. The creation operator $\ad_{R+}$ creates 
a right handed gluon traveling in the positive $z$ direction. The remaining creation 
operators have obvious meanings. The out state is the photon at rest and thus has the rotation 
properties of the spherical harmonics $Y_{1,m}(\h,\phi)$. As the in state has both gluons 
either right handed or left handed, the $z$ component of the net angular momentum is zero.
Hence photon out state also has $m$ = 0.

We will rotate the in state and the out state 
by angle $\pi$ about the $z$-axis and then about the $x$-axis.
The photon out state, mimicking the rotation properties of $Y_{1,0}(\h,\phi)$, 
is an eigenstate of either rotation with eigenvalues $+1$ and $-1$ respectively.
Focusing on the in state, we  note that rotation 
by an angle $\phi$ about the axis $\hat{n}$
is achieved by the action of the appropriate operator $U(R^n_\phi)$ on the state in question, 

\bea
U(R^z_\phi) |  R+ ; R+ \rc &=& U(R^z_\phi) \ad_{R +} \ad_{R +}  |0 \rc \nn \\
&=&  U(R^z_\phi)  \ad_{R +}  U^{-1}(R^z_\phi) U(R^z_\phi)  \ad_{R +}  U^{-1}(R^z_\phi) 
U(R^z_\phi) | 0 \rc.
\eea

\nt Recalling the action of the rotation operators on creation operators (see Ref. 
\cite{wei95}), we obtain, 

\bea
U(R^z_\phi)  \ad_{R +}  U^{-1}(R^z_\phi)  = \sum_{h} \mathcal{D}(R^z_\phi)_{R h}  \ad_{h,\hat{p}}. 
\label{genrot}
\eea

\nt Where $\mathcal{D}(R^z_\phi)_{R h}  =  \lc R | e^{ i J_z \phi } | h \rc$ 
is the rotation matrix for the rotation of the state (in this case vector). 
The index $h$ runs over all the possible 
$z$ components of the spin of the particle. The vector $\hat{p}$ represents the new direction 
of motion of the particle after rotation. The action of any unitary operator, such as a rotation, 
on the vacuum will result in the vacuum again. Setting $\phi = \pi$ we obtain the simple relation 
for the 
action of the rotation 
operator on the gluon creation operator,

\bea 
U(R^z_\pi)  \ad_{R +} U^{-1}(R^z_\pi) &=& e^{ i \pi} \ad_{R +}  \nn \\
U(R^z_\pi)  \ad_{R -} U^{-1}(R^z_\pi) &=& e^{ - i \pi} \ad_{R -}
\eea 

\nt Using the above it is not difficult to demonstrate that the in state of two gluons is 
an eigenstate of  $R^z_\phi$ with eigenvalue +1. Thus, both in state and out state are 
eigenstates of $R^z_\phi$ with the same eigenvalue. As a result, there is no restriction 
to this transition, on the basis of this symmetry. 

We now concentrate on rotation by $\pi$ about the $x$ axis. 
The outstate is an eigenstate of this operation with eigenvalue -1.
Using Eq. (\ref{genrot}) we note that,

\bea
U(R^x_\pi)  \ad_{R +}  U^{-1}(R^x_\pi)  &=&  \ad_{R -} \nn \\
U(R^x_\pi)  \ad_{R -}  U^{-1}(R^x_\pi)  &=&  \ad_{R +} \nn \\
U(R^x_\pi)  \ad_{L +}  U^{-1}(R^x_\pi)  &=&  \ad_{L -} \nn \\
U(R^x_\pi)  \ad_{L -}  U^{-1}(R^x_\pi)  &=&  \ad_{L +} 
\eea

\nt One may, thus, demonstrate that the two gluon 
in state is an eigenstate of the above rotation with eigenvalue +1, 

\bea 
U(R^x_\pi) \left( | R+ ; R- \rc - | L+ ; L- \rc \right) 
&=&   \left[ U(R^x_\pi) \ad_{R+} U^{-1}(R^x_\pi) 
U(R^x_\pi)\ad_{R-} U^{-1}(R^z_\phi) \right.  \nn \\ 
&-&  \left. 
U(R^x_\pi) \ad_{L+} U^{-1}(R^z_\phi)  U(R^x_\pi) \ad_{L-}  U^{-1}(R^z_\phi) \right] 
U(R^x_\pi) | 0 \rc \nn \\
&=& \left[ \ad_{R-} \ad_{R+} - \ad_{L-} \ad_{L+} \right] | 0 \rc \nn \\
&=& | R- ; R+ \rc - | L- ; L+ \rc. 
\eea
This implies that this 
transition is not allowed by any interaction. Thus, we demonstrate 
Yang's theorem in the vacuum: this transition is not allowed


\subsection{Yang's theorem in media}


The above argument for no transition has been formulated for two 
massless vectors fusing to a 
spin one final state in the vacuum. We now intend to extend 
this to a transition in the medium. 
One may argue at this point that the correct method of analyzing this 
situation would be to 
start from a particular many body state; invoke the matrix 
element of the transition (this would give us the requisite 
creation and annihilation operators) 
and end up in a particular final many body state

\bea
\mathcal{M} = \lc n_1^f , n_2^f ... n_{\infty}^f |  \left( \int d^4 x H_{I} (x)
\right)^n | n_1^i , n_2^i
... n_{\infty}^i \rc 
\eea
 
This has to be followed by squaring the matrix element and weighting it 
by the Boltzmann factor $e^{-\B E_i}$, where $E_i$ is the total energy of the
in state, $\B$ is the inverse temperature. Then this quantity must be summed over
all initial and final states to obtain the total transition probability per unit 
phase space for this process as

\bea
\mathcal{P} = \sum_f \sum_i e^{ - \B E_i } \Bigg| \lc n_1^f , n_2^f ... n_{\infty}^f 
|  \left( \int d^4 x H_{I} (x) \right)^n | n_1^i , n_2^i
... n_{\infty}^i \rc  \Bigg|^2     
\eea

\nt The above method though comprehensive, does not allow a simple
amplitude analysis as the case for the vacuum. Such an analysis
may be constructed by drawing on the spectator analysis of loop 
diagrams. This method as applicable to this process has been expounded in 
the previous section. Our results are essentially contained in Fig. \ref{spec2}

The following analysis with spectators may appear to be rather heuristic at 
times. The reader not interested in such a discussion may consider the 
fact that the introduction of the medium formally involves the introduction 
of a new four-vector ${\bf n}$ into the problem. If we were to consider 
the case of dileptons produced back-to-back in the rest frame of the 
medium, the  
results from the vacuum should still hold as in this case 
the only new ingredient is a new four-vector of the bath $({\bf n} = (1,0,0,0))$. This 
four-vector is obviously rotationally invariant and cannot in any way introduce 
rotational non-invariance via dot or cross products with any three-vector in the 
problem. However, if the two gluons are not exactly back-to-back or equivalently 
the medium has a net three-momentum, then rotational invariance is explicitly 
broken. Even if we were to boost to the frame where the gluons are 
exactly back-to-back, we would find the medium streaming across the reaction. 
The above argument for the validity of the theorem for static dileptons will 
now be demonstrated via the spectator interpretation.

We consider the Feynman diagrams of Figs. \ref{spec} and \ref{spec2}.
The effect of the medium, on the transition, is understood as a change in
the in state to include an incoming quark from a particular 
quantum state $\sg$. Where, 
the index $\sg$ will be used to indicate all the 
characteristics of the quark in question 
such as the momenta, spin or helicity, colour etc. 
The out state 
will also be modified as indicated to include a quark 
emanating from the transition and 
re-entering the medium in the same quantum state $\sg$ vacated by 
the incoming extra quark. In the discussion 
that follows, we will keep 
referring to the original state containing the two incoming gluons as the 
in state, and the one with the outgoing 
dilepton as the out state. The extra particles that enter the reaction 
from the medium or exit the reaction and go back into the 
medium will be referred to 
as `medium particles'.
The full effect of the medium will only be incorporated on 
summation of  the transition 
rates obtained by including all such states $\sg$ 
weighting the entire process (incoming particles $\ra$ reaction $\ra$ outgoing particles) 
by appropriate thermal distribution factors for the incoming and outgoing medium particles.
The thermal factors will essentially be those of Eq. (\ref{spec_eqn}). 
No doubt, there must also appear thermal factors for the incoming gluons.
For the duration of the entire discussion, 
we will constrain the two gluons to have the same momenta; 
the distribution functions will thus play no role, and hence have been ignored.

The new total in states and out states will now be given by state vectors that look like,
\[
\sum_\sg \Bigg( | R+ ; R- \rc | \sg \rc - | L+ ; L- \rc | \sg \rc   \ra  
| \g^* \rc | \sg \rc  \Bigg) \Big[ 1 - 2 \nf(E_\sg) \Big] .
\]
 
\nt In the above equation, we have taken the incoming and outgoing particle from the medium to 
be a fermion, as is appropriate in this case. The reader will recognize the
thermal factors to be exactly those of Eq. (\ref{mtrx_elm}). 
Each state may once again be obtained by the 
action of the corresponding creation operators on the vacuum state. The new additional 
factors $\nf(E_\sg)$ are the appropriate distribution functions, used in the expressions 
to indicate particles leaving and entering the medium. 
Unlike the in and out states, the contributions from these 
medium states are added coherently, \ie one does not square the amplitude and 
then sum over spins and momenta but rather the procedure is carried out in
reverse, as indicated by Fig. \ref{spec2}. 
The sum $\sum_\sg$, represents integration over all momenta, sum over spins and colours etc.

Our method of extending Yang's symmetry will involve identifying certain subsets 
of the entire sum to be performed, which will turn out to be eigenstates of the 
rotation and parity operations to be carried out once more on these states. The 
argument will essentially be the following: if we can decompose the entire in and 
out state into certain subsets, with each subset being an eigenstate of the 
symmetry operator with the same eigenvalue, then the entire in and out states 
will also be eigenstates with the same eigenvalues. Then, as for the vacuum 
process, we will compare the eigenvalues for the in state and outstate.

To illustrate, we focus on a subset of four terms in the full sum in which 
one of the incoming medium-fermions has a three-momentum $\vq$. To keep 
the discussion simple we pick $\vq$ to be in the $yz$ plane (the discussion 
may be easily generalized to include $\vq$ in an arbitrary direction). The 
four processes under consideration are:

\bea
\Big[ \left(| R+ ; R- \rc  - | L+ ; L- \rc \right)  | \vq ; \ua \rc   &\ra & 
| \g^* \rc | \vq ; \ua \rc \Big] (1 - 2 \nf(E_{\vq,\ua}) )   \label{eigenx} \\
\mbox{} + \Big[ \left(| R+ ; R- \rc  - | L+ ; L- \rc \right)  | \Ro^z_{\pi} \vq ; \ua \rc   &\ra & 
| \g^* \rc |\Ro^z_{\pi} \vq ; \ua \rc  \Big]  (1 - 2 \nf(E_{\Ro^z_{\pi} \vq,\ua}) )  \nn \\
\mbox{} + \Big[ \left(| R+ ; R- \rc  - | L+ ; L- \rc \right)  | \Ro^x_{\pi} \vq ; \da \rc   &\ra & 
| \g^* \rc |\Ro^x_{\pi} \vq ; \da \rc \Big]   (1 - 2 \nf(E_{\Ro^x_{\pi} \vq,\da}) )  \nn \\
\mbox{} + \Big[ \left(| R+ ; R- \rc  - | L+ ; L- \rc \right)  
|\Ro^z_{\pi} \Ro^x_{\pi} \vq ; \da \rc   &\ra & 
| \g^* \rc | \Ro^z_{\pi}\Ro^x_{\pi} \vq ; \da \rc  \Big]  
(1 - 2 \nf(E_{\Ro^z_{\pi}\Ro^x_{\pi} \vq,\da}) ). 
\nn
\eea

\nt Where, $\Ro^z_{\pi} \vq$ represents the three-momentum $\vq$ rotated by an angle $\pi$
about the $z$ axis, $\Ro^x_{\pi} \vq$ represents $\vq$ rotated about by an angle $\pi$ about the 
$x$ axis. The arrows $\ua, \da$ represent the $z$ component of the spin of the medium-fermion.  
As we are in the centre of mass of the thermal bath we have

\bea
E_{\vq,\ua} = E_{\Ro^z_{\pi} \vq,\ua} = E_{\Ro^x_{\pi} \vq,\da} = E_{\Ro^z_{\pi}\Ro^x_{\pi} \vq,\da}.
\label{engsame}
\eea

\nt Thus we may completely factor out the distribution functions. 
Without loss of generality we 
may combine all four in states and out states to give,

\bea
& & \left(| R+ ; R- \rc  - | L+ ; L- \rc \right)  \llb | \vq ; \ua \rc  + | \Ro^z_{\pi} \vq ; \ua \rc 
+  | \Ro^x_{\pi} \vq ; \da \rc  +  |\Ro^z_{\pi} \Ro^x_{\pi} \vq ; \da \rc \lrb \nn \\
&\ra&  | \g^* \rc   \llb | \vq ; \ua \rc  + | \Ro^z_{\pi} \vq ; \ua \rc 
+  | \Ro^x_{\pi} \vq ; \da \rc  +  |\Ro^z_{\pi} \Ro^x_{\pi} \vq ; \da \rc \lrb 
\label{thermal_yang}
\eea

\nt Now, it is simple to demonstrate using the methods of rotation of creation 
operators outlined in the vacuum case, that both the in and out states are eigenstates 
of $\Ro^x_{\pi}$. Concentrating on the rotation of the in state we 
obtain 

\bea 
& & U(R^x_\pi) \left(| R+ ; R- \rc  - | L+ ; L- \rc \right)  
\llb | \vq ; \ua \rc  + | \Ro^z_{\pi} \vq ; \ua \rc 
+  | \Ro^x_{\pi} \vq ; \da \rc  +  |\Ro^z_{\pi} \Ro^x_{\pi} \vq ; \da \rc \lrb \nn \\
&=& U(R^x_\pi) \left( \ad_{R;+} \ad_{R;-}  - \ad_{L;+} \ad_{L;-} \right)  U^{-1}(R^x_\pi)  \nn \\
& & \llb U(R^x_\pi) \ad_{\vq ; \ua}  U^{-1}(R^x_\pi) 
U(R^x_\pi) \ad_{\Ro^z_{\pi} \vq ; \ua}  U^{-1}(R^x_\pi)
U(R^x_\pi) \ad_{\Ro^x_{\pi} \vq ; \da}  U^{-1}(R^x_\pi)
U(R^x_\pi) \ad_{\Ro^z_{\pi} \Ro^x_{\pi} \vq ; \da}  U^{-1}(R^x_\pi) \lrb | 0 \rangle \nn \\
&=& -i \left(| R+ ; R- \rc  - | L+ ; L- \rc \right) 
\llb | \Ro^x_{\pi} \vq ; \da \rc  + |\Ro^x_{\pi} \Ro^z_{\pi} \vq ; \da \rc
+  | \vq ; \ua \rc  +   | \Ro^z_{\pi} \vq ; \ua \rc \lrb  
\label{thermal_yang2}
\eea

\nt Note that the medium-fermions just mix into each other, but the over all state remains 
the same. Following the above method one can show that the outstate is also an 
eigenstate of $\Ro^x_{\pi}$ but with an eigenvalue of $i$. Thus, we can decompose the 
entire sum over spins and integration over the three-momenta of the medium-fermions 
into sets of states as indicated, each will result in an in state and an out state 
between which no transition is allowed. For the rotation $\Ro^z_{\pi}$ we note that 
the eigenstates are in fact a subset of two states: in this case, the sum of the 
first two states of Eq. (\ref{eigenx}) are eigenstates of $\Ro^z_{\pi}$; as is 
the sum of the third and fourth state. 

This would imply that such a transition, as implied by the Feynman diagrams of 
Fig. \ref{spec}, can not occur. There is however a caveat to the above discussion. 
Note that in the vacuum case we expressly boosted to the frame where the two 
gluons would be exactly back-to-back with their three-momenta equal and opposite. 
Then, rotational symmetry was invoked to demonstrate the impossibility of this 
transition. In the case of the processes 
occurring in medium, we tacitly began the analysis with the two gluons once 
again exactly back-to-back in the rest frame of the bath. 
However, if the two gluons are not exactly back-to-back or equivalently 
the medium has a net three-momentum, then rotational invariance is explicitly 
broken. Even if we were to boost to the frame where the gluons are 
exactly back-to-back, we would find the medium streaming across the reaction.
This would make the distribution functions of the two gluons different (even though in this 
frame they have the same energy), Eq. (\ref{engsame}) would no longer hold. As a 
result it will not be possible to construct eigenstates of the rotation 
operators $\Ro^z_{\pi}$ and $\Ro^z_{\pi}$ as done previously. As the in and out 
states will no longer be eigenstates of $\Ro^z_{\pi}$ and $\Ro^z_{\pi}$ with 
different eigenvalues, transitions will, now, be allowed between them.

In the above discussion, we have demonstrated how the medium may, once again, 
break another symmetry of the vacuum; in this case rotational symmetry. 
This allows the transition of Fig. \ref{2vert} to take place in the medium.
This process is strictly forbidden, in the vacuum, by two different 
symmetries (charge conjugation and rotation). It is forbidden in the exact back to 
back case by rotational symmetry in a $C$ broken medium \ie the effect is zero
for $\vp=0$. To obtain a non-zero contribution, rotational 
symmetry has to be broken by a net $\vp$. The magnitude of the 
signal from such a symmetry breaking effect may only be deduced via 
detailed calculation. In the next section we shall outline just such a 
calculation.

The derivation of Yang's theorem depended expressly on the the two incoming 
gluons being massless. This enforced their polarizations to be purely
transverse. Another way of breaking rotational invariance is thus by giving 
masses to the gluons. This is not unjustified since in medium they acquire a 
thermal mass. 
If the gluons are considered as being massive, this implies that they now have three rather 
than two physical polarizations. The longitudinal polarization state  
is then physical and can be seen as being responsible for the breaking 
of Yang's theorem. Under parity, rotation around the $x$--axis of $\pi$, 
and rotation around the $z$--axis of $\pi$, the creation operator of 
the new $0$--polarization state transforms, respectively, as
\begin{eqnarray}
\mathcal{P}a^{\dagger}_{0\pm}\mathcal{P}^{-1} = -a^{\dagger}_{0\mp} \\
U(R^x_\pi)a^{\dagger}_{0\pm}U^{-1}(R^x_\pi) = -a^{\dagger}_{0\mp}\\
U(R^z_\pi)a^{\dagger}_{0\pm}U^{-1}(R^z_\pi) = a^{\dagger}_{0\pm} 
\end{eqnarray}
Then we can construct the two new in states with total angular 
momentum along the z-direction of $+\hbar$:
\begin{equation}
\frac{1}{\sqrt{2}}\left\{|R+;0-\big> -|0+;L-\big>\right\}
\end{equation}
and  $-\hbar$:
\begin{equation}
\frac{1}{\sqrt{2}}\left\{|0+;R-\big> -|L+;0-\big>\right\}
\end{equation}
We must now inquire as to the possibility of a transition 
with a virtual final photon in the $m=\pm1$ states. First 
note that now the out states are not eigenstates of $U(R^x_\pi)$ 
operator, but still are of $U(R^z_\pi)$ with eigenvalue $-1$. 
Applying this operator on the in states, one finds that they are 
eigenvectors with eigenvalue $-1$. Therefore, the in states and 
out states share the same eigenvalues. Thus the transition is not 
prohibited by parity and rotational symmetries.

In a real medium both effects discussed 
(\ie finite momentum and massive gluons) 
would be present and simultaneously lead to 
the breaking of this symmetry. In this article 
we separate these two effects and study each 
in turn. The complete calculation incorporating 
the two simultaneously will be left for a future effort \cite{maj04}.

There remains yet another means by which the symmetry of Yang's 
theorem may be broken: that of a rotationally non-invariant regulator. 
The results from this scenario have already been presented in Refs. \cite{maj01,maj01b}.
In these calculations the dileptons were produced back-to-back, \ie  from a virtual 
photon which is static in the rest frame of the plasma from the fusion of 
massless gluons. The reader will note that Yang's theorem holds for 
such a process and hence should result in a vanishing rate. Yet this rate was
found to be non-zero. 
The reason behind this result is the choice of the regulator used in those 
calculations. From Eqs. (\ref{thermal_yang}) and (\ref{thermal_yang2}) we note that 
we required the coherent sum of at least four quark states, with the incoming 
quark occupying symmetric angles, for Yang's theorem to hold. If we designate 
one of the gluons to be along the $z$ axis, and one of the incoming quarks 
is assigned the momenta ($q,\h,\phi$), then the configuration that obeys the 
rotational symmetry of Yang's theorem will include incoming quarks at 
($q,\h,\phi+\pi$), ($q,\pi - \h, \phi$) and ($q,\pi-\h, \phi + \pi$). 
Thus in the $\h$ integration, one must include balancing contributions 
$\h$ and $\pi-\h$. 

As is the case in this article, the results of Refs. \cite{maj01,maj01b} 
consisted of the sum of contributions from multiple residues, some 
of which displayed singularities as $\h \ra 0$ or $\h \ra \pi$.
This corresponds to the one of the internal lines in Fig. \ref{spec} going 
on shell. This divergence is canceled when all the different residues are
combined, as will be shown in the next section. Each residue is 
then evaluated using a regulator. Two obvious choices are the angle 
$\delta < \h < \pi - \delta$, and the magnitude of three-momentum of the intermediate 
state $ x = | \vq + \vk | = \sqrt{q^2 + k^2 + 2kq\cos{\h}}$ where $|q - k| + \e < x < q+k - \e$. 
All residues are then evaluated in the limit $\kd \ra 0$ or $\e \ra 0$ where 
the same regulator is used throughout. The divergence will be canceled 
in either case when all the residues are summed. In the calculations of 
Refs. \cite{maj01,maj01b} the three-momentum regulator was chosen. 

From the preceding discussion, it is obvious that integration using 
the angular regulator $\kd$ obeys Yang's theorem as symmetric contributions 
from $\h = \kd$ and $\h= \pi - \kd$ are included. However this is not the case
with the three-momentum regulator. Though  $\e = 0 $ corresponds to $\kd=0$, at 
$\e \ra 0$ we find that $x= |q + k| - \e $ corresponds to a $\h = \kd_1 \ra 0$, 
while $x= |q - k| + \e $  corresponds to a $\h = \pi - \kd_2 \ra 0$.
After some calculation, it may be demonstrated that 

\bea 
\kd_1 \simeq \sqrt{ \frac{2 |k+q| \e}{kq}} 
\eea

\nt while

\bea 
\kd_2 \simeq \sqrt{ \frac{2 |k-q| \e}{kq}} 
\eea
 
\nt Thus for a given $k,q$ the same $\e$ corresponds to different limits of 
the angular integration. The rotational invariance required for Yang's 
theorem is broken and a non-vanishing contribution results. The physical interpretation 
of the non-vanishing results of Refs. \cite{maj01,maj01b} thus become unclear.
The results obtained here do not depend on those prior findings. In this sense,
the current article constitutes an update and a correction.
In what follows, the symmetry in Yang's 
theorem will be broken only by real physical effects prevalent in hot media.

\section{Results for $\vp \neq 0$ }

We concentrate on the breaking of the symmetry underwriting Yang's theorem by imposing 
a non-zero three-momentum to the process. In other words, the virtual photon has a 
net three-momentum in the rest frame of the bath. The gluons are considered to 
be massless. As may be easily understood from the preceding section, the
magnitude of symmetry breaking rises with the magnitude of the three-momentum. 
Thus the largest possible values of $\vp$ will lead to the largest signals.

We want in the end to calculate the number of dileptons per unit 
spacetime per unit energy per unit $|\vp|=p$, i.e., 

\[
\frac{d^6N}{d x^4  dE dp} = \frac{d^2R}{dE dp} = \int d^2 \Ow _p p^2  \frac{d^4R}{dp^4}
\]

\nt Where we have integrated the differential rate over all solid angles
$\Ow_p$. The angle of $\vp$ is always measured from the direction of the 
more energetic incoming parton: in this case the gluon with energy $>E/2$. 
This procedure will also be followed for the Born term and will be
explained in greater detail in the last subsection.

We begin by presenting results for a simpler case. We look at the differential 
rate when the two incoming quarks or gluons are 
forced to be back-to-back but have different energies.
This may be obtained by setting the angle $\kd$ between the photon and the in
coming gluon to zero. Alternatively one may obtain this by 
expanding the rate in a Taylor expansion in angle and
keeping only the first term. One reason for considering this special case is
that this is the simplest generalization from the $\vp=0$ case. 
Yet another reason for considering this case is the possibility of 
an analytical solution. We provide complete analytic results in this case,
as opposed to the general case where the final integrations can only be
performed numerically.


\subsection{Differential rate for back-to-back gluons}


We begin by evaluating the  sum of the six matrix elements on the 
right hand side of Eq. (\ref{spec_eqn_b}). Recall that $i,j$ indicate 
the polarizations of the incoming gluons where as $l$ is the polarization 
of the outgoing photon.  In this configuration, a variety  of simplifying
relations result:

\begin{equation}
\mat_{+,-,l} = \mat_{-,+,l} = 0
\end{equation}

\nt as expected (and pointed out before), in a back-to-back situation, 
both gluons have to arrive with the 
same polarization, \ie either both must be right handed or both left handed.
This will result in a net $z$ component of angular momentum $L_z = 0$.
The configuration with one left-handed and one right-handed in a 
back-to-back scenario will have an $L_z = 2$ and thus will not couple 
to a spin one object. 

\begin{equation}
\mat_{i,j,+} = \mat_{i,j,-} = 0
\end{equation}
 
\nt There is no contribution to the transverse modes of the photon. This
is obvious as the two gluons form a system with a net $L_z = 0$.

We are now in a position to write down explicit expressions for 
the various matrix elements of Eq. (\ref{spec_eqn_b}), \ie  
$\mat_{i,j,l} =  \left( \mps_{i,j,l}^1 + \mps_{i,j,l}^2 + 
\mps_{i,j,l}^3 + \mps_{i,j,l}^4 + \mps_{i,j,l}^5  + \mps_{i,j,l}^6 \right) \kD n (E_q,\mu)$. 
On performing all the angular integrations,
frequency sums and contractions with polarization vectors, the results are:

\bea 
\mbox{} \!\!\!\!\!\!\!\!\!\!\!\!\!\!\!\!\!\!\!\!\!\!\!\!\!\!\!\!\!\! \mat_{+,+,3} 
&=& \mat_{-,-,3}  
=\frac{ e g^2 \kd^{bc} }{2} \int  \frac{dq q^2 \D \nf (E_q, \mu)}{(2\pi)^3 E_q \sqrt{E^2-p^2} } \Bigg[
32\,{\frac {\pi \,{m}^{2} p \left (4\,{q}^{2}-{p}^{2}+{E}^{2}+4\,{m}^{2}
\right )\ln (\left |-\sqrt {{q}^{2}+{m}^{2}}+q\right |)}
{\left[ (E-p)^2 
- 4(q^2 + m^2) \right] \left[ (E+p)^2 - 4(q^2+m^2) \right] q }} \nn \\
&-& 32\,{\frac {\pi \,{m}^{2}p\left 
(4\,{q}^{2}-{p}^{2}+{E}^{2}+4\,{m}^{2}\right )\ln (\sqrt {{q}^{2}+{m}^
{2}}+q)}
{\left[ (E-p)^2 - 4( q^2 + m^2 ) \right] 
\left[ (E+p)^2 - 4( q^2 + m^2 ) \right] q }} \nn \\
&-& 16\,{\frac {
\pi \,{m}^{2}\left (E-2\,\sqrt {{q}^{2}+{m}^{2}}\right )\ln (\left |-1
/2\,{E}^{2}+1/2\,{p}^{2}+E\sqrt {{q}^{2}+{m}^{2}}-qp\right |)}{q\left 
(E+p-2\,\sqrt {{q}^{2}+{m}^{2}}\right )\left (2\,\sqrt {{q}^{2}+{m}^{2
}}-E+p\right )}} \nn \\
&+& 16\,{\frac {\pi \,{m}^{2}\left (2\,\sqrt {{q}^{2}+{m}
^{2}}+E\right )\ln (\left |-1/2\,{E}^{2}+1/2\,{p}^{2}-E\sqrt {{q}^{2}+
{m}^{2}}-qp\right |)}{q\left (-2\,\sqrt {{q}^{2}+{m}^{2}}-E+p\right )
\left (2\,\sqrt {{q}^{2}+{m}^{2}}+E+p\right )}} \nn \\
&+& 16\,{\frac {\pi \,{m}^
{2}\left (E-2\,\sqrt {{q}^{2}+{m}^{2}}\right )\ln (\left |-1/2\,{E}^{2
}+1/2\,{p}^{2}+E\sqrt {{q}^{2}+{m}^{2}}+qp\right |)}{q\left (E+p-2\,
\sqrt {{q}^{2}+{m}^{2}}\right )\left (2\,\sqrt {{q}^{2}+{m}^{2}}-E+p
\right )}} \nn \\
&-& 16\,{\frac {\pi \,{m}^{2}\left (2\,\sqrt {{q}^{2}+{m}^{2}}+
E\right )\ln (\left |-1/2\,{E}^{2}+1/2\,{p}^{2}-E\sqrt {{q}^{2}+{m}^{2
}}+qp\right |)}{q\left (-2\,\sqrt {{q}^{2}+{m}^{2}}-E+p\right )\left (
2\,\sqrt {{q}^{2}+{m}^{2}}+E+p\right )}}  \Bigg] \nn \\  \label{anal_sum}
\eea

We are thus concentrating on the 
virtual photon produced only by back-to-back gluons of unequal 
momenta, and will compare with the rate of 
production from only back-to-back quarks of unequal momenta. 
We are thus breaking Yang's symmetry by the introduction of 
a net three-momentum $p$. 
It is a simple exercise to check that the above matrix element 
vanishes linearly with $p$ as $p \ra 0$. The apparent pole in $q$, 
is canceled between the six terms.
There is still the $dq$ integration to be 
performed, this is done numerically.   

The differential production rate for pairs of massless leptons with total
energy $E$ and and total momentum $\vec{p}$ 
is given in terms of the discontinuity in the
photon self-energy as (see Eq. (\ref{r}))

\begin{equation}
\frac{dR}{ d^4 p  } = \frac{ e^2 }{ 3 ( 2 \pi )^5 }
\frac{ r }{ (\fp^2) } 
\frac{1}{ e^{\B E} - 1 } 
\end{equation}

\noindent Where 
$r = \left[ \frac{ p_\mu p_\nu }{ \fp^2 } - 
g_{ \mu \nu } \right] \dsc \left[ -i \Pi^{ \mu \nu } \right] $.   
In general, the matrix element depends on the angle $\kd$ between $\vp$ and $\vk$. As a
result, two equivalent means of angular integration present themselves: we
may set $\vp$ along the $z$ direction; in which case the matrix element will
depend on the polar angle $\kd = \theta_k$ of $\vk$. As a result, the integral over
$\Ow_p$ 
yields an overall factor of $2 \pi$. Alternatively we may set $\vk$
along the $z$ direction, in which case the rate depends on the angle $\kd =
\theta_p$, and the integral over $\Ow_k$ yields an overall factor of $2 \pi$.
Both methods are equivalent. We chose the latter prescription. 
Thus we calculate the derivative of the differential rate
with respect to the incoming gluon angle $\Ow_k$

\begin{equation}
\frac{dR}{ d^4 p d \Ow_k } = \frac{ e^2 }{ 3 ( 2 \pi )^5 
(\fp^2)  \left[ e^{\B E} - 1  \right] } 
\frac{ d r }{ d \Ow_k } . \label{difdifrate}
\end{equation}

\begin{figure}[htbp]
\hspace{-1.5cm}
\resizebox{2.75in}{2.75in}{\includegraphics[0in,1in][6in,9in]{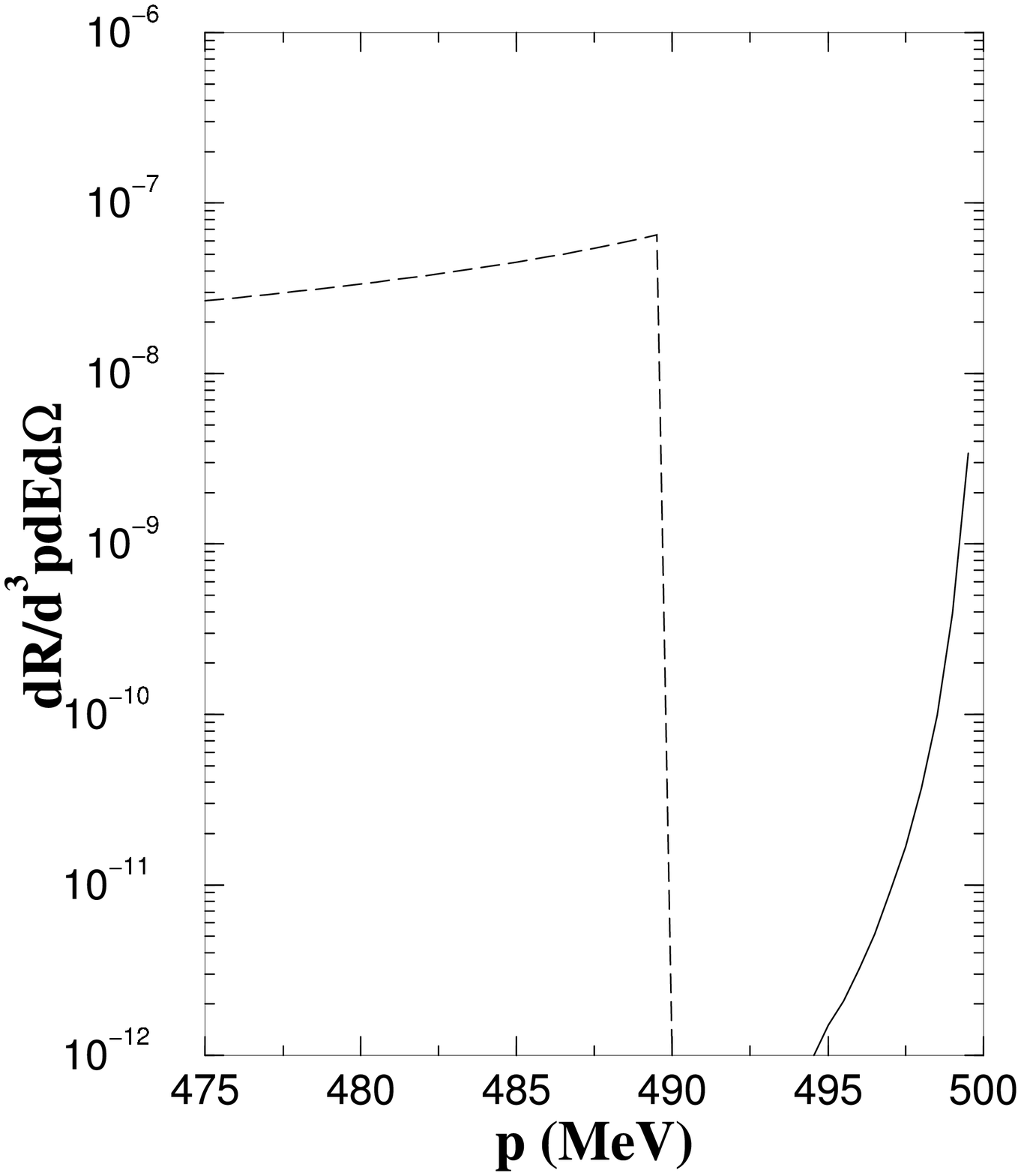}}
\hspace{1.5cm}
\resizebox{2.75in}{2.75in}{\includegraphics[0in,1in][6in,9in]{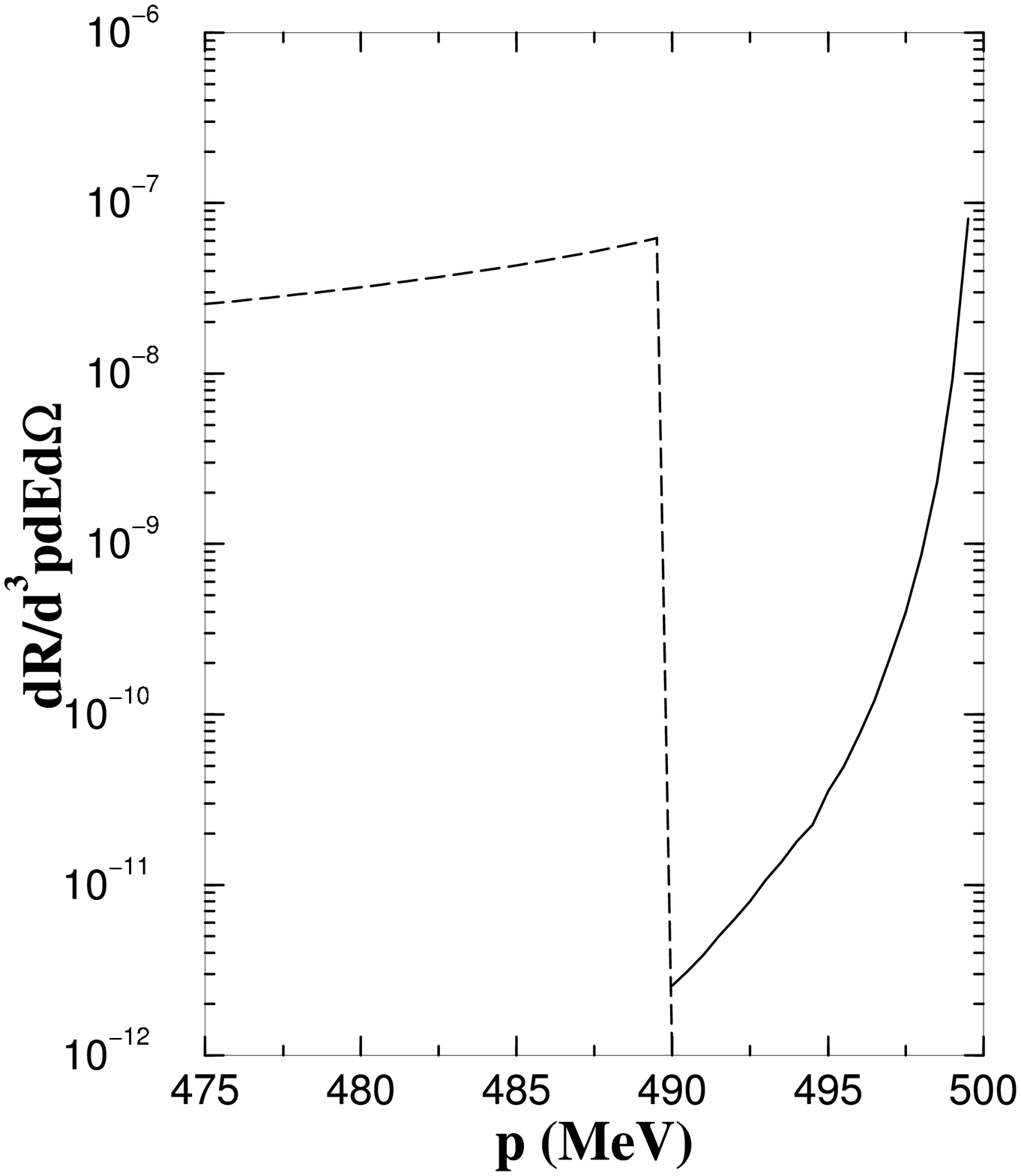}}
\caption{The differential production rate of low mass dileptons
from two back-to-back processes. Invariant mass runs from 156 MeV to 0 MeV. The energy 
of the dilepton is $E=500$MeV, and the abscissa is the three-momentum $p$.
The dashed line represents the contribution from the process 
$q\bar{q}\rightarrow e^{+}e^{-}$. The solid line corresponds to 
the process $gg\rightarrow e^{+}e^{-}$. Temperature is 400 MeV. 
Quark chemical potential is 0.1T. The second figure is the same as the first but with
 $\mu=0.5T$  }
    \label{nmu}
\end{figure}

As mentioned before, temperatures in the plasma formed at RHIC and LHC 
have been predicted to lie in the range from 300-800 MeV \cite{wan96,rap01}.
 For this calculation,  
we use  $T=400\mbox{ MeV and } 800 \mbox{ MeV}$. To evaluate the effect
 of a finite chemical potential we perform the calculation  
with two values of
chemical potential $\mu=0.1T$ (left plot in Fig. \ref{nmu}) and 
$\mu=0.5T$ (Right plot in Fig. \ref{nmu})
\cite{gei93}.
This calculation is performed for two flavours of quarks  
with current masses.

\begin{figure}[htbp]
\hspace{-1cm}
 \resizebox{2.75in}{2.75in}{\includegraphics[0in,1in][6in,9in]{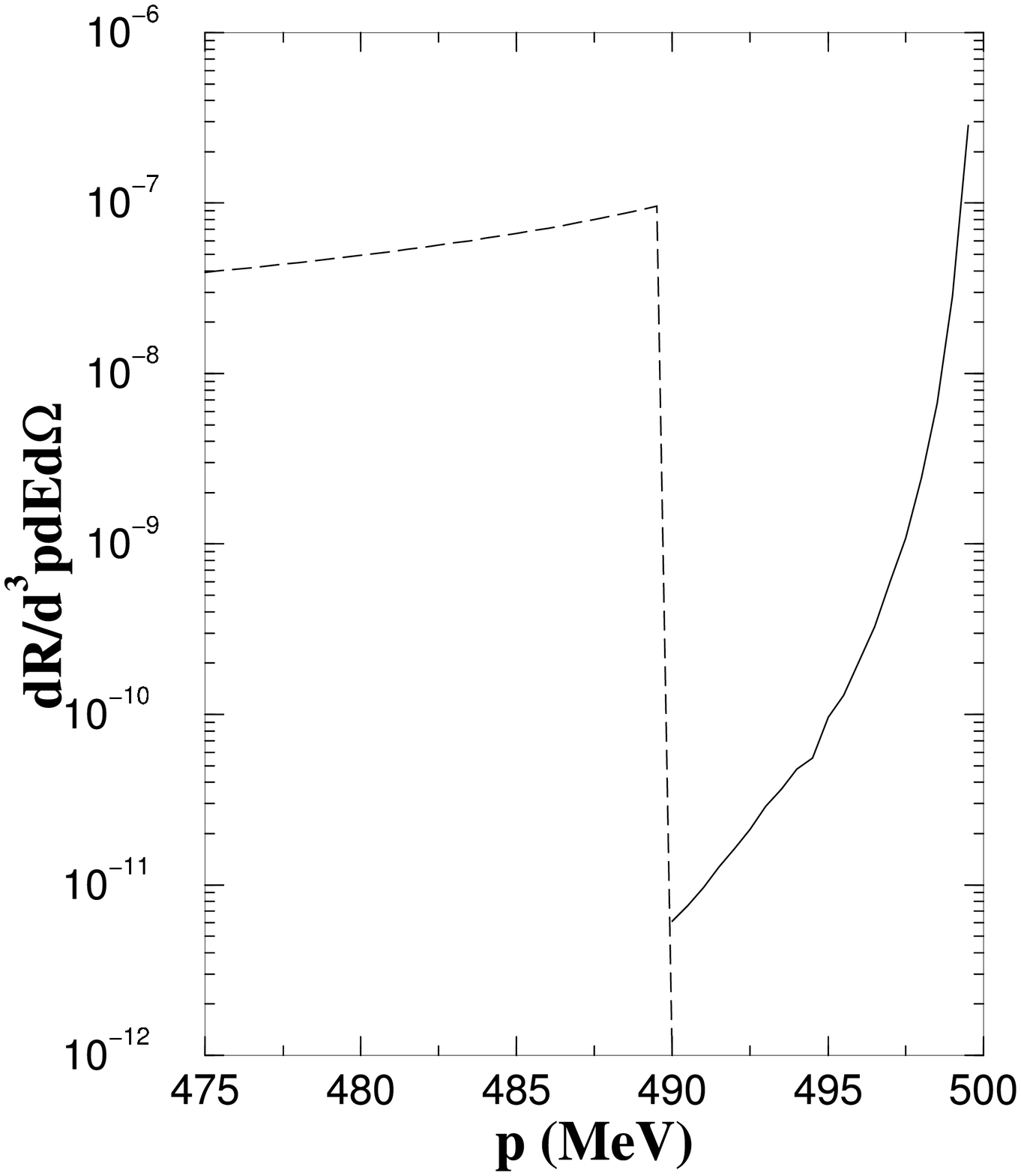}}
\hspace{1.5cm}
 \resizebox{2.75in}{2.75in}{\includegraphics[0in,1in][6in,9in]{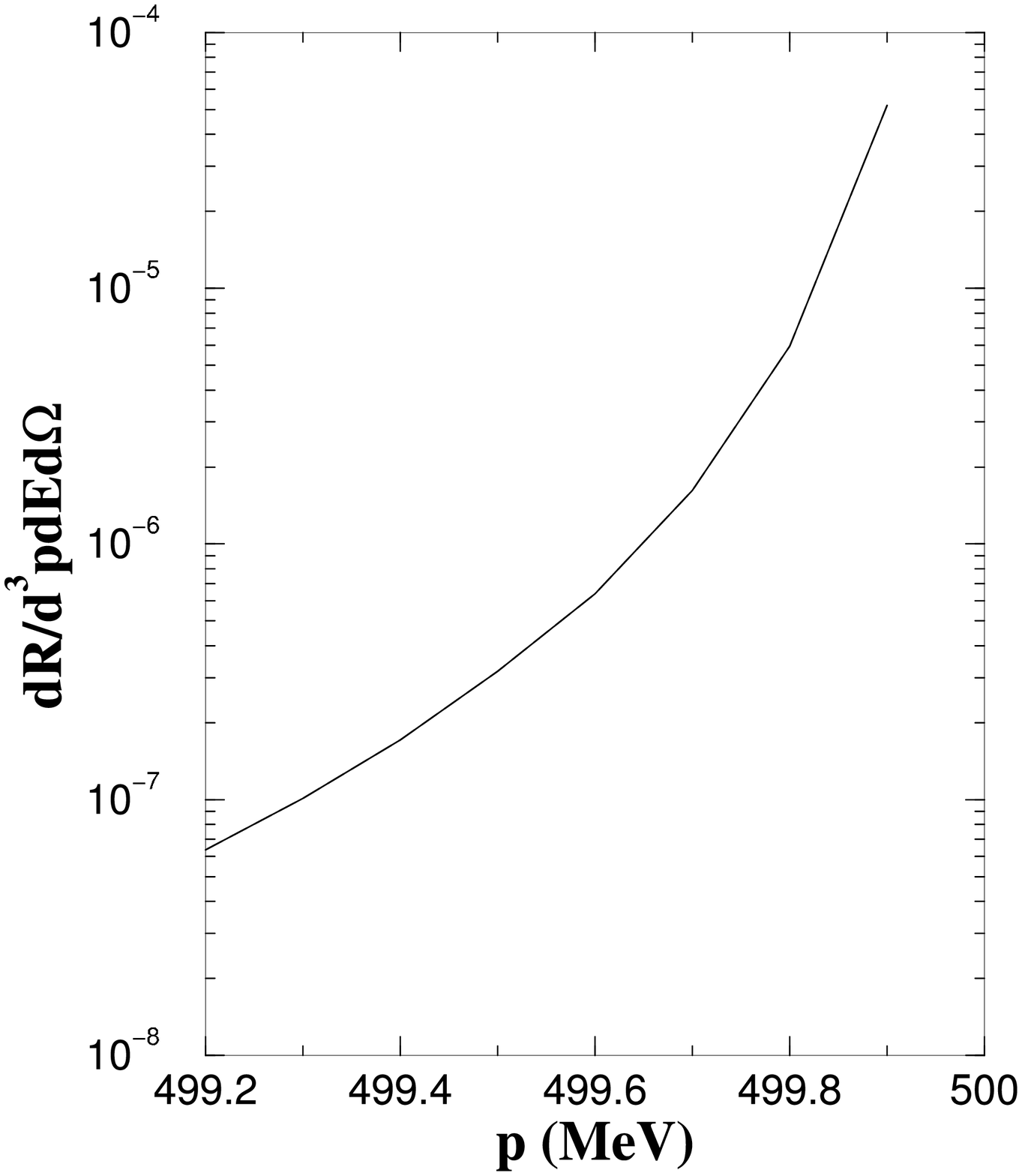}}
    \caption{Left panel is same as Fig. \ref{nmu} but with a temperature of 800 MeV. 
Quark chemical potential is 0.5T. Right panel is the rate of $gg \ra e^+ e^-$ just below and beyond the 
Born term threshold. Born term threshold is at $p=499.6$ MeV for a dilepton energy of $E=500$ MeV and 
quark current masses of $10$ MeV.}
\label{hitemp}
\end{figure}

 In Fig. \ref{nmu}, 
the differential rate (Eq. (\ref{difdifrate})) 
for the production of dileptons with an 
invariant mass from 0 to 156 MeV is presented. 
The energy is held fixed at 500 MeV and the three-momentum $p$ of 
the dilepton is varied (a dilepton invariant mass $M=156$ MeV for an 
energy $E=500$ MeV corresponds to the three-momentum of the 
dilepton $p=475$ MeV). 
In the figures, the dashed line is the rate from tree level
 $q\bar{q}$ ; the solid line is that from the process 
 $gg\rightarrow e^{+}e^{-}$.  
We note that in both cases the gluon-gluon process dominates
 at very low mass and dies out at higher mass leaving the $q\bar{q}$ 
process dominant at higher mass or lower momentum. The Born 
term displays a sharp cutoff at photon invariant mass $ M = \sqrt{2mE} $. 
The back-to-back 
annihilation of two massive quarks (of mass $m$) 
to form the virtual photon of energy $E$ and invariant mass 
$M$ is no longer 
kinematically allowed.
Also, the annihilation of a quark antiquark pair to form a 
dilepton is not allowed for any incoming angle for dileptons 
with an invariant mass $M < 2m$. The gluons 
being massless, continue to contribute in this region: this 
contribution is shown in the right 
panel of Fig. \ref{hitemp}. Thus the signal from $gg$ fusion (for partons with current masses) 
is dominant at low invariant masses 
for intermediate dilepton energies. In Fig. \ref{hitemp}, we indicate the influence of 
a higher plasma temperature on the rates. Here, a plasma 
temperature of 800 MeV and 
$\mu=0.5T$ is used; the left panel displays the rates 
below the Born term threshold and the right panel  
displays the rates above threshold. We note in the left panel of 
Fig. \ref{hitemp}, as expected, that the gluon fusion term rises further 
due to thermal loop enhancement, however this rise is rather minimal.  
In the right panel we note that the rates for $gg \ra e^+ e^-$ continue to 
rise due to the growing distribution functions for soft gluons. In a realistic 
calculation the gluons would be endowed with a thermal mass that would cut off 
the steep rise in the $gg$ rate.
 
\subsection{Full differential rates}


In the previous subsection, we calculated the six matrix elements of
Eq. (\ref{spec_eqn_b}) (or Fig. \ref{spec}), for the special 
case of the angle between $\vk$ 
and $\vp$ \ie $\kd = 0$. A close inspection of the diagrams of
Fig. \ref{spec} will convince the reader that, in the event that the in
coming gluons and the outgoing photon are in the same line, the angular
integration over all quark directions displays an azimuthal symmetry. This
implies that the integration $d \phi_q$ results in a mere overall factor of
$2\pi$. 
The $d\h$ integration, though non-trivial, can be performed analytically 
and results in Eq. (\ref{anal_sum}). The integration over the quark magnitude 
cannot be performed analytically and we resorted to numerical means. 

In the case of a $\kd \neq 0$, the azimuthal symmetry in the $d^3 q$ integration
is absent and both $d \phi_q$ and $d \h_q$ integrations are non
trivial. It is no longer possible to perform both analytically. Following the $dq$
integration, we also have to perform the
integration over the angle $\kd$. This will give us the
differential rate $\frac{d^2 R}{dEdp}$. This turns out to be a complicated 
problem to solve in general. However, from the previous subsection we 
have learnt that the rate form this process is comparable to the Born term only
at very low mass $M = \sqrt{E^2 - p^2} \ra 0$, or rather $p \ra  E$ (see
Figs. \ref{nmu},\ref{hitemp}). If we insist on calculating solely in this
limit an approximation scheme may be constructed. 

There are two basic scales that we input into this problem: the mass of the
quarks $m$ and the temperature of the plasma $T$ (the chemical potential is 
always estimated as a fraction of the temperature: hence it does not constitute a 
separate scale). At the low invariant masses
(of observed dileptons) in question the strange quark does not contribute. 
For the up and the 
down quark, we are considering plasmas where $m<<T$. We now insist on observing
dileptons with large four-momenta $E,p \sim T$, yet very small invariant mass $M =
\sqrt{E^2 -p^2} \sim m << T$. Yet another smaller scale is that of $x  =  E -p$
where $x = \frac{M^2}{E + p} \sim \frac{M^2}{E} <<  M $. 
One may construct three dimensionless scales
from these quantities: $ 1 >> \frac{M}{E} >> \frac{x}{E} $.

We denote the incoming gluons by their polarizations $i,j$. Now, say $i$ is
more energetic and is ascribed the momenta $k$, the other ($j$) has momentum
$|\vp - \vk|$ by conservation. 
As outlined in the previous section, we intend to
integrate $\kd$ from $\kd=0$ when the two gluons are back-to-back and 
$k >> | \vp - \vk | = E - k$ ($i$ is a very hard gluon and $j$ is very soft) up to 
$\kd = \kd_{max} $ where $k = | \vp - \vk | = E/2$ 
(where, throughout the gluon
$i$  is more energetic than the gluon $j$). The remainder of
the $\kd$ integration may be obtained by simply replacing $i$ with $j$ 
and noting that the remainder is nothing but the same integration with
the gluon $j$ now ascribed the larger energy $k$ and $\kd$ defined such that 
the $j$ points in the positive $z$ direction. The magnitude of $\kd_{max}$
may be estimated simply from the preceding discussion. As the gluons are
massless

\bea
E - k = E_{| \vp - \vk |} = | \vp - \vk | = \sqrt{ p^2 + k^2 - 2 p k \cos(\kd) }
\eea

\nt This implies that 

\bea
 \cos (\kd) = \frac{ 2 E k - M^2  }{ 2 p k }   
\eea

\nt The value of $\kd = \kd_{max}$ occurs at $k = E/2$, hence

\bea
\cos ( \kd_{max} )  &=&   1  -  \frac{ \kd^2}{2}  =  \frac{p}{E}  \nn \\
&=& 1 - \frac{x}{E} 
\eea
 
\nt Thus  $\kd_{max} = \sqrt{\frac{2x}{E}} \sim \frac{M}{E} $. 

Yet another inference about the behaviour of the rate with $\kd$ may be drawn
from Eq. (\ref{r}). Here we note the presence of two Bose-Einstein distribution
functions: $n(k) , n(E-k)$ and the factor $\frac{k}{E-k}$ in the measure. Note 
that as $\kd \ra 0$, $k$ tends to its maximum value, and $\w = E-k$ 
tends to its minimum value. This greatly enhances the factor 
$\frac{k}{E-k} n(k)n(E-k)$, as compared to its value at $\kd = \kd_{max}$, 
where $k$ and $E-k$ are of the same magnitude. This 
implies that if the the matrix element does not rise sharply with $\kd$ then 
the differential rate falls off as $\kd$ is raised.

The above mentioned observations allow us to expand the matrix element in a
series in $\kd$. On expansion we note the following behaviour for the
longitudinal photon :

\[
\mat_{i,j,3}(\kd) = m_0  -  m_2 \kd^2 + ...
\]

\begin{figure}[htb]
\hspace{-3cm}
\resizebox{2.75in}{2.75in}{\includegraphics[0in,1in][6in,9in]{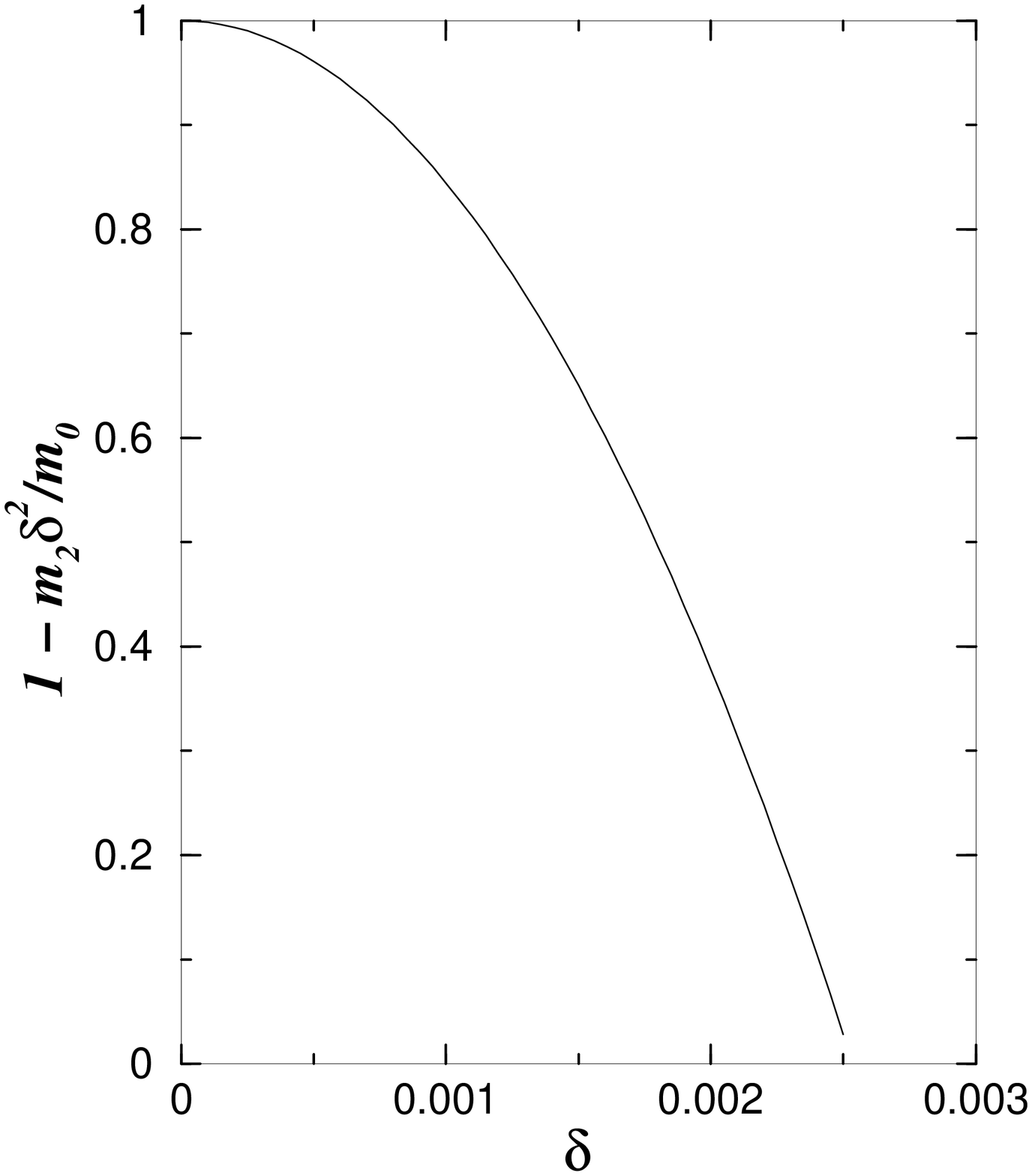}} 
\hspace{2cm}
\resizebox{2.75in}{2.75in}{\includegraphics[0in,1in][6in,9in]{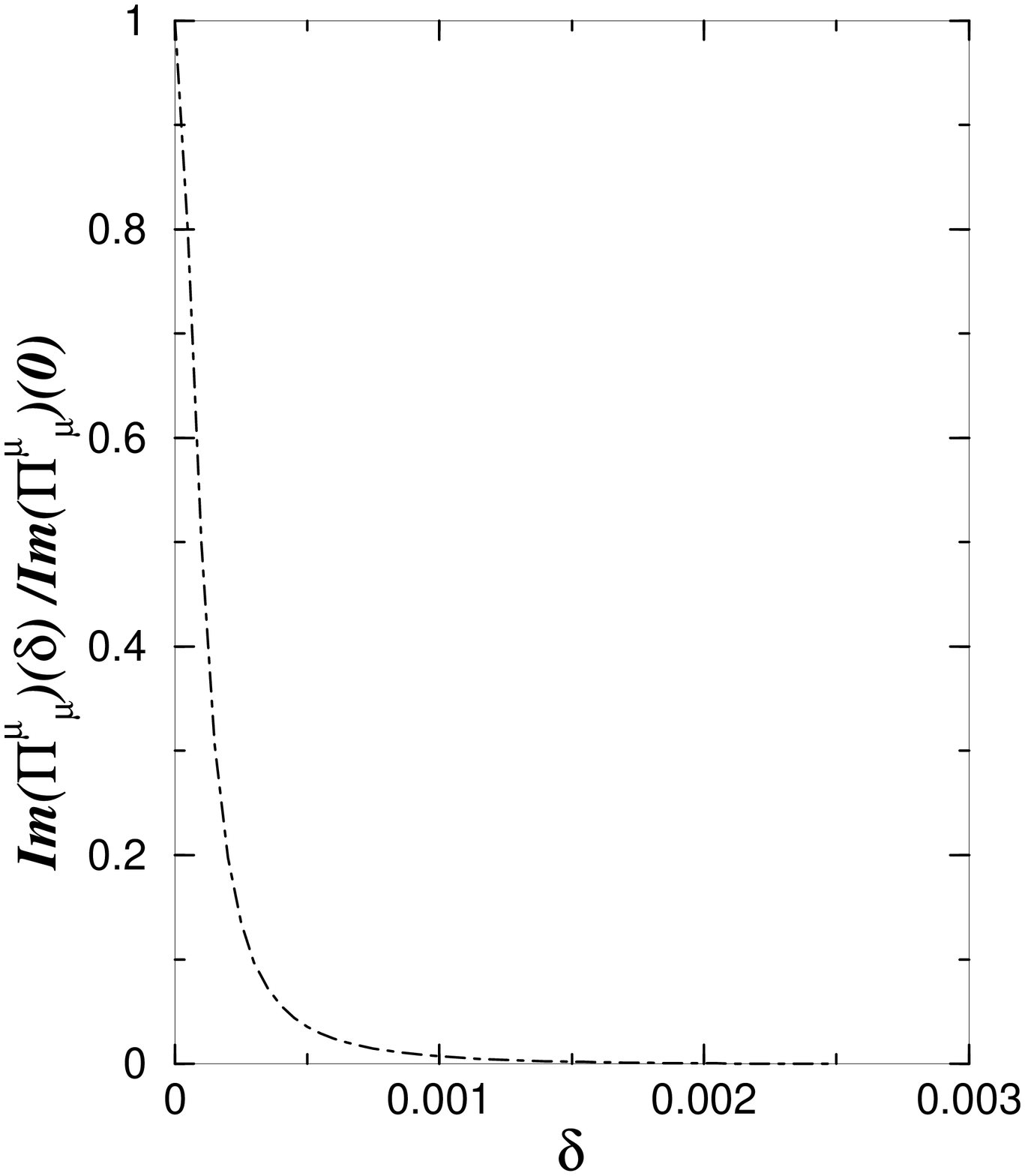}} 
\caption{Left panel shows the behaviour of $ (M = m_0  -  m_2 \kd^2)
/M(\kd = 0) $ as a function of $\kd$ for a typical case of $T=0.5$GeV,
$\mu= 0.5T$, $E=1$GeV and $p=0.9999$GeV.  Right panel shows the behaviour
of $ Im(\Pi^{\mu}_{\mu})/Im(\Pi^{\mu}_{\mu})(\kd = 0) $ as a
function of $\kd$. }
\label{maple_1}
\end{figure}

\nt Where the $m_i$'s are all positive contributions, and depend on $E,p,T,\mu$. 
Thus for small $\kd$ the matrix element (or the square of the matrix element) 
drops as $\kd$ rises from zero. A plot of this behaviour 
(in arbitrary units) for a typical case is
shown in Fig. \ref{maple_1}. It should be pointed out 
that the angular integrations of the 
quark momenta (\ie $\h, \phi$) may be analytically 
performed only after the 
expansion in $\kd$. The remaining integration over the 
quark momentum $q$ is performed numerically. 
If the matrix element were not expanded in a series in 
$\kd$ one would have to perform four sets 
of integrations numerically. The presence of poles 
in the matrix elements of the diagrams in 
Fig. \ref{spec} makes this a prohibitively difficult procedure.   

Now that we have moved away form the 
back-to-back scenario, we will also witness the production of transverse photons. 
On expansion in 
$\kd$ we note the following behaviour:

\[
\mat_{i,j,\pm}(\kd) = m_1 \kd  -  m_3 \kd^3 + ...
\]

\nt As expected this contribution goes to zero as $\kd \ra 0$. It also turns
out to very much smaller than the longitudinal contribution in the limit of
small invariant mass dileptons. We thus obtain that the dominant contribution 
to the rate emanates from the longitudinal photons.  

\begin{figure}[htbp]
 \rotatebox{-90}
{\resizebox{5in}{5in}{\includegraphics[0in,2in][8in,10in]{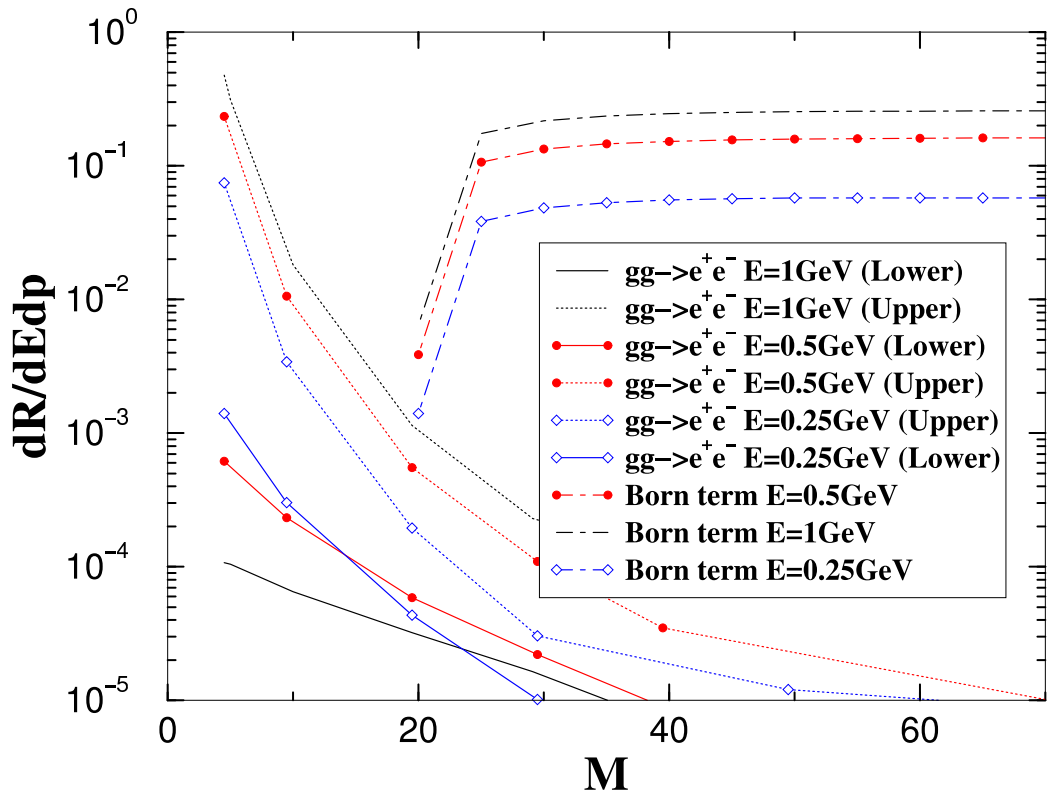}}} 
    \caption{(Color online) Differential rate for the production of dileptons at high
    momenta and small invariant mass. See text for details. }
\label{fin1}
\end{figure}

It should be pointed out
that such an expansion is only strictly valid as long as $\kd$ is of the order
of the smallest scale in the problem \ie $\frac{x}{E}$. However, as demonstrated
by the plot of the imaginary part of the self energy in Fig. \ref{maple_1}, 
it remains valid much beyond this point. This is due to the
influence of the measure and the gluon distribution functions which 
drop rapidly as one moves away from $\kd= 0$. It should be pointed out that
only  the matrix elements have been expanded in a series, all other factors
(\eg gluon distribution functions, measures \etc) retain their closed expressions in $\kd$.

We are now in a position to integrate over $\kd$. This is also performed
numerically. As we have expanded the matrix elements in $\kd$
and retained only a finite number of terms, the square of the matrix elements
$|\mat |^2$  grow beyond a certain $\kd = \kd_1$. 
This growth is not real and is merely a facet of our finite
expansion in $\kd$. Two possible means of carrying out the $\kd$ integration 
present themselves. We may terminate the integral at $\kd_1$. 
Ostensibly, this represents the lower limit of the rate. These are 
represented in Fig. \ref{fin1} as the solid lines (both with and without symbols).
We may 
continue to integrate up to $\kd = \kd_{max}$; this will include the integration 
of a growing rate convoluted with an increasing angular measure ($\sin\kd$). This 
represents the upper limit of the rate. These are represented in Fig. \ref{fin1} as the 
dotted lines. As the invariant mass is lowered or the 
energy of the pair is raised the differential rate drops sharply from its value at
$\kd=0$; this invariably results in $m_2 >> m_1$. As a result the integration  
beyond $\kd = \kd_1$ produces a large contribution; this in turn leads to the excess 
growth displayed by the upper limit at small invariant mass and large energies.

The rates after integration over $\kd$ are presented in
Fig. \ref{fin1}. The dot-dashed lines are the rates from the Born term at various 
energies of the virtual photon. The
solid lines are the lower limits of the rates from gluon gluon fusion for the 
respective energies of the virtual photon. 
The dotted lines are the upper limits of the respective rates. 
In general the rates from gluon gluon
fusion are suppressed as compared to the Born 
term except at very high momenta or very low invariant
mass. 
Three cases have been presented where the energy 
of the dilepton is set at 0.25, 0.5 and 1GeV. 
As may be noted, the results are quite similar 
to the back-to-back gluon fusion rates, \ie the 
case at $\kd=0$. The rate from gluon fusion 
rises beyond the Born threshold due to the 
rising Bose-Einstein distributions of soft 
massless gluons.


\section{Results for $m_g > 0$}


We now turn to the case where gluons acquire a medium-induced mass 
and where the virtual photon has no net three-momentum ($\vec{p} = 0$). 
As we have seen, the gluon's longitudinal degree of polarization 
allows to circumvent Yang's theorem. More specifically, 
we expect transitions 
to occur when the net component of angular momentum is $\pm \hbar$:   

\begin{eqnarray}
\mathcal{M}_{i,j,3} &=& 0 \\
\mathcal{M}_{\pm,3,\pm} &=& \mathcal{M}_{3,\pm,\pm} = 
-\frac{m_g}{k}\frac{ e g^2 \kd^{bc} }{2}\int 
\frac{dq}{(2\pi)^2} \Delta \widetilde{n}(E_q,\mu)J(q,k)
\end{eqnarray}

\nt where

\begin{equation}
J(q,k) = \frac{qk+\frac{q}{k}m_g^2}{8E_kE_q}\ln\left[\frac{(E_qE_k+
qk)^2-\frac{m_g^4}{4}}{(E_qE_k-qk)^2-\frac{m_g^4}{4}}\right]
-\frac{q}{4k}\ln\left[\frac{E_q^2E_k^2-
\left(qk-\frac{m_g^2}{2}\right)^2}{E_q^2E_k^2-
\left(qk+\frac{m_g^2}{2}\right)^2}\right]
\end{equation}

\nt where $E_k =\sqrt{k^2+m_g^2}$. Therefore, in the 
back--to--back configuration with massive and 
equally energetic gluons we find that the non-zero 
matrix elements are proportional to the gluon mass 
and scale like $\sim \mu$ confirming the finite density 
nature of this process. We point out that the integrand 
diverges unless $m_g < 2m_q$. Beyond this limit the 
self-energy analytical structure becomes intricate. 
In this section, we present results only where the 
above relation holds. Dileptons production rates 
from regions beyond this threshold, as well as rates 
emanating from the fusion of massive gluons with 
$\vec{p} \neq 0$ will be addressed in a future effort \cite{maj04}.

\begin{figure}
\begin{center}
\includegraphics[width=15cm]{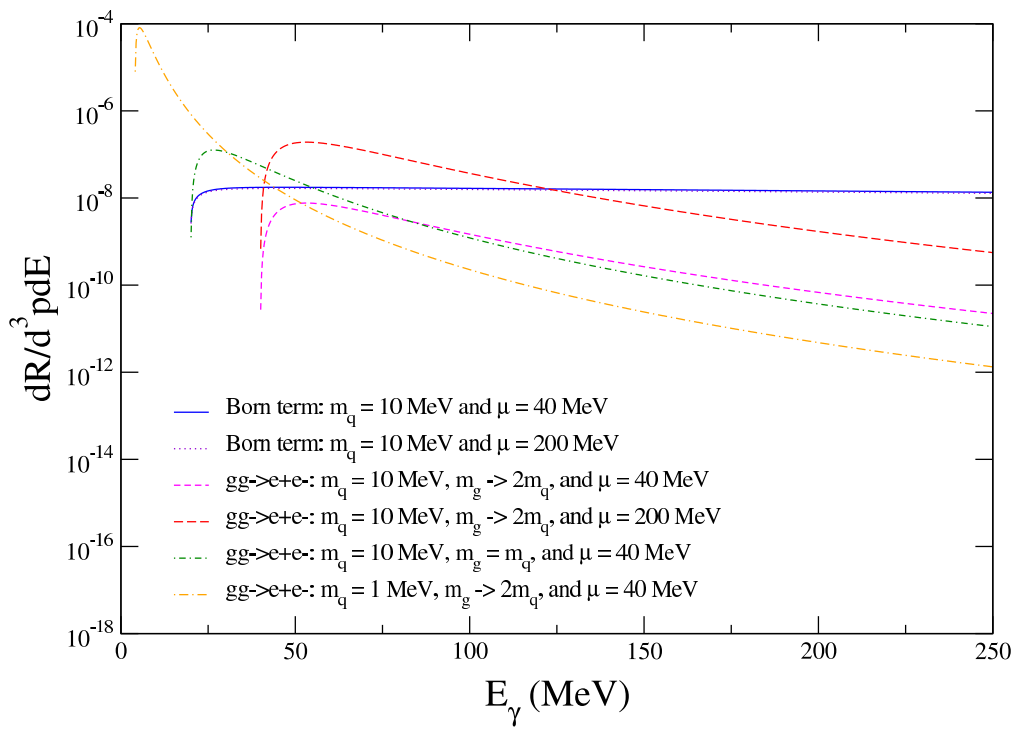}
\caption{(Color online) Differential production rates for $T=400$ MeV. See text for details}
\label{fig1}
\end{center}
\end{figure}

\begin{figure}
\begin{center}
\includegraphics[width=12cm,angle=0]{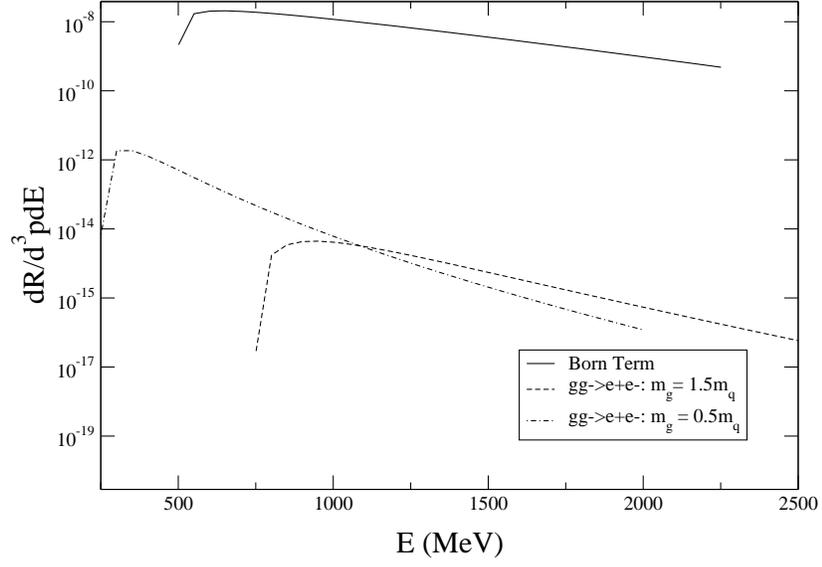}
\caption{Differential production rates for $2g \rightarrow l\bar l$ 
with $T=400$ MeV, $\mu_q=40$ MeV, $m_q=250$ MeV, and different gluon masses.}
\label{fig2}
\end{center}
\end{figure}

To explore dilepton production in the range where the mass 
inequality is respected, we begin by ascribing 
current masses of 10 MeV to the quarks, setting the quark 
chemical potential to $\mu = 40$ MeV, and the gluon mass to almost twice 
that of the quark (\ie $m_g=19.99$ MeV) as a reference point. 
With these, we see that the differential production rate 
due to the gluon fusion is lower than the contribution 
from the Born term across the range of invariant mass 
(see Fig.~\ref{fig1}). However, if we increase the 
quark chemical potential to $\mu = 200$ MeV or reduce the 
quark current mass to $m_q = 1$ MeV while maintaining 
the gluon to quark mass ratio (\ie 1.999), we see that 
the gluon fusion rate may dominate over the Born term 
up to an invariant mass of 125 MeV in the case where 
the quark chemical potential reaches 200 MeV. We also present 
results where 
the gluon mass is set equal to that of the quark (i.e. 10 MeV), in this case 
we see that the rate from the gluon fusion dominates 
at low invariant mass. If we lower the gluon 
mass beyond that of the quark then the process will 
have a non-vanishing contribution in a region forbidden 
for the $q\bar q$ process due to its threshold (see Fig.~\ref{fig2}).

If instead of current masses, the quark masses are set 
to values of the order of $gT$ then we find the rate 
from gluon fusion to be suppressed as compared to 
the Born term (see Fig.~\ref{fig2}). This is not 
unexpected as the gluons fuse through a quark triangle; 
the presence of large masses in the quark propagators 
leads to the rates being suppressed in this region of parameter space.
 
\begin{figure}
\begin{center}
\includegraphics[width=12cm]{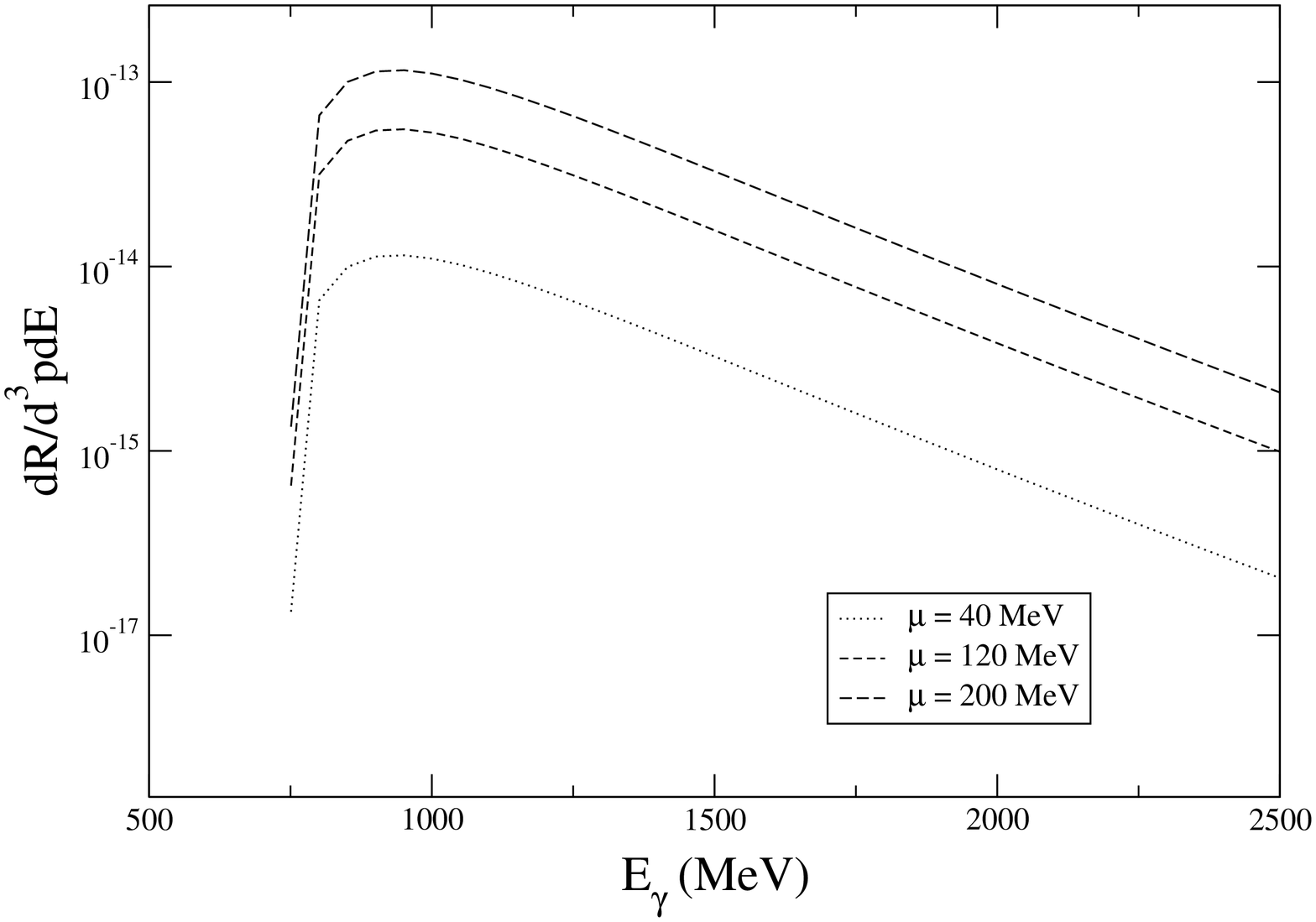}
\caption{$\mu^2$ scaling of the differential production 
rates with $T=400$ MeV, $m_q=250$ MeV and $m_g=1.5m_q$ MeV.}
\label{fig3}
\end{center}
\end{figure}

As in the case of gluon fusion with $\vec{p} \neq 0$, 
an interesting feature of the differential rate is 
its strong dependence on $\mu^2$ (Fig.~\ref{fig3}) 
and its weak dependence on temperature and other 
energy scales: as the chemical potential increases, 
the rate rises. For the Born term the opposite 
behaviour is true. As the chemical potential increases, 
the antiquark population is depleted, inhibiting the 
production of dileptons through this channel. Thus, 
an accurate estimate of the differential rate will 
require a good knowledge of the baryon chemical potential 
as well as its variation with time in a QGP.

\section{Discussions and Conclusions}


In this article we have presented a detailed study of 
the observational effects of broken charge conjugation, 
and broken rotational invariance in a QGP formed 
in a heavy-ion collision. 
The signal under consideration was the spectrum of 
dileptons emanating from such a medium. The reason 
behind this choice is evident: electromagnetic 
signatures provide a direct probe of all time sectors of a heavy-ion 
collision. The breaking of these symmetries is 
manifested at lowest order in the spectrum of 
dileptons produced by two gluon fusion into a virtual photon 
through a quark loop.

Such a process is forbidden in the vacuum by both 
Furry's theorem and Yang's theorem. 
Charge conjugation was broken explicitly
by the introduction of a non-zero population of 
$u$ and $d$ valence quarks in the plasma. 
A non-zero baryon density, present solely in these 
flavours causes a net electric charge density in 
the medium. This leads to the breaking of Furry's theorem 
which holds purely in neutral media. 

The presence of a preferred rest frame of the medium leads to the 
introduction of a bath four-vector {\bf n} into the 
problem. If calculations are performed in this frame 
then {\bf n}$=(1,0,0,0)$. No such vector exists in the vacuum. If two massless 
back-to-back gluons with equal energy fuse in vacuum along
the $z$-axis through the quark triangle, then 
the out state consisting of a static virtual photon will 
have the $z$ component of its spin $J_{z} = 0$. 
Both in state and out state are eigenstates of a rotation by $\pi$
about the $x$-axis but have different eigenvalues. Hence such a transition 
is forbidden, this is the statement of Yang's theorem and is based on the
invariance of both in state and out state under a rotation by 
$\pi$ about the $x$-axis. 

The above argument of no transition (due to the in state and out state 
possessing different eigenvalues with respect to the rotation by $\pi$ 
about the $x$-axis) continues to hold even for the production of a static photon 
in the rest frame of the bath where 
the in state and out state are modified to include spin-half fermions.
It should be pointed out that we have tacitly assumed the plasma to 
be infinite in extent and isotropic; a realistic 
plasma of finite extent which is not 
spherically symmetric will explicitly break rotational invariance. 
To our knowledge, this fact may be understood solely in the spectator 
interpretation of loop diagrams.
The spectator interpretation represents a formal procedure by which the 
imaginary part of a diagram containing loops may be re-expressed in terms of the 
product of matrix elements consisting solely of tree diagrams and  particles 
from the bath that do not partake in the reaction process. The spectator 
interpretation for the imaginary part of a three-loop diagram was derived 
for this process and is essentially contained in Fig. \ref{spec2}. In the 
spectator interpretation different states containing fermions with different spins 
are added coherently. This allowed us to construct a subset of the entire 
sum of in states that respected the rotation symmetry of the vacuum state 
(see Eq. (\ref{thermal_yang})). For each such subset no transition was 
allowed by arguments similar to those used in the construction of Yang's 
theorem (see Sec. IV B). 

This invariance will be broken 
by the presence of any three vector in the problem. The two possible 
choices for such a three-vector are a net three-momentum of the virtual photon 
($\vp \neq 0$ in the bath frame), 
or a non-zero $z$-component of its spin ($J_z \neq 0$).
In the first case one may boost to the rest frame of the static 
photon. However, this will no longer be the rest frame of the plasma 
and rotational invariance will be explicitly broken. 
In the second case the production of a virtual photon with a 
$J_z \neq 0$ will require an incoming massive gluon which 
breaks one of the principal conditions required for 
Yang's theorem to hold.
In a real QGP we would expect both effects to be present simultaneously. 
In the interest of a clearer understanding of the mechanism of 
symmetry breaking we had chosen to explore both possibilities in 
isolation. 
  
In the case of a virtual photon with a net three-momentum $\vp$, we began 
with the case of two massless gluons in a back-to-back configuration 
along the $z$-axis but with unequal energies. 
This resulted in a virtual photon with a net three-momentum along 
the $z$-axis. Under this kinematic restriction we 
present a closed analytical expression for the matrix element 
(see Eq. (\ref{anal_sum})). The rates from this matrix element are 
plotted in Figs. \ref{nmu} and \ref{hitemp} in comparison with the Born term. These 
figures represent the cases with plasma temperatures at 400 MeV and 800 MeV, 
corresponding to the cases of QGP formed at RHIC and LHC energies. 
We find as expected that the differential rate rises 
with increasing chemical potential $\mu$; two extreme cases with $\mu=0.1T$ 
and $\mu=0.5T$ have been explored. 
The rates however show a rather modest rise with increasing temperature. 
This has been pointed out earlier due to the vanishing of the $g^2 T^2$ 
component in the HTL calculation of this loop diagram \cite{maj01b}.
The rates are also observed to rise with increasing three-momentum $|\vp|$ 
(for a given fixed energy) as expected due to the breaking of Yang's theorem. As the gluons are 
massless they continue to contribute into the region beyond the Born threshold 
(see Fig. \ref{hitemp}).
In this region, contributions originate from the fusion of a gluon carrying 
a large majority of the photons energy with an ultra-soft gluon carrying a 
tiny fraction of the photon energy. 
In fact the rising rate in the right panel of Fig. \ref{hitemp} is due to 
the Bose enhancement obtained from the distribution function of the soft gluon.
The presence of a gluon dispersion relation or a gluon mass will lead to this 
rate reaching a maximum at a threshold set by twice the gluon mass. 

Integrating over all incoming gluon angles turned out to be an involved 
procedure and led us to invoke an approximation scheme. At very small invariant
mass, we noted that the gluon fusion rate is dominated by back-to-back gluon fusion. We 
thus expanded in the angle between the photon and the more energetic gluon $\delta$.
Results for the differential rate per unit energy and per unit momentum $d^2R/dEdp$ 
have been presented in Fig. \ref{fin1}, in comparison with the Born term. 
As expected the rate from gluon-gluon fusion 
becomes comparable to the Born term  only at very low invariant mass, for photon energies 
of 0.25, 0.5 and 1GeV. 
As noted in the previous section a large portion of the
enhancement may be attributed solely to the lack of a mass for the gluons and 
Bose-Einstein distributions as opposed to a Fermi-Dirac distributions for the quarks.

In the case of a virtual photon with a net $J_z$, the possible choices are 
$J_z = \pm 1$. This requires one of the incoming gluons to be in a 
longitudinally polarized state. Hence, the gluons were endowed with a mass. 
One may ascribe the origin of such a mass to dispersion in the medium. 
As in the previous case, the rate is seen to rise sharply with increasing 
chemical potential. Due to analytic considerations, the quark mass ($m_q$)
was always set to be larger than half the gluon mass ($m_g < 2 m_q$). 
In this kinematic region, the rates from gluon gluon fusion 
turned out to dominate over the Born term for low  
invariant masses of dileptons, if the quarks were chosen to be light $m_q << T$. 
However, the rates were sub-dominant to the Born term for 
the production of dileptons with large invariant mass, or for quark masses $\sim gT$.
We point out, that, in this calculation $T = 400 MeV$, hence, $g \sim 2$. Thus 
unlike the case for plasmas at very high temperature $gT \sim T$.   

Gluons with masses at and above this threshold ($2m_q$) 
may decay into two quarks which are both 
simultaneously on shell. In the language of spectators, this corresponds to 
one of the propagators in the diagrams of Fig. \ref{spec} going on shell.   
The calculational and interpretive complications that arise from 
this situation are rather 
involved and represent a problem for the spectator interpretation. 
A preliminary calculation 
without the use of the spectator interpretation at this threshold in the 
limit of massless quarks and gluons found the 
rate to be large \cite{maj01}; however these required the use of 
momentum cutoffs which did not respect the 
symmetry required by Yang's theorem. As pointed 
out earlier this made the physical interpretation of 
these results unclear.  
The computation of the rates at and beyond this threshold 
will full quark and gluon dispersion relations at finite 
three-momentum 
and their interpretation in terms of the spectator picture 
will be dealt with in a subsequent calculation \cite{maj04}. 
Our goals in the present article have been to separately elucidate certain 
symmetries of the vacuum which are broken by a particular channel of 
dilepton production in a quark gluon plasma. For the 
purposes of simplicity we computed the rates from this channel 
for a plasma in complete thermal and chemical equilibrium. 
As the dileptons in this channel are produced essentially form 
the fusion of gluons, this process may display far greater 
significance in early plasmas which are estimated to be out of chemical 
equilibrium with large gluon populations. An accurate extimation 
of the full dilepton rate from this 
channel will require a two step process. One needs to combine the 
two effects of symmetry breaking discussed in this article. 
This rate will then have to be folded in with a realistic space-time 
model of the evolution of the plasma. Such a model will 
have to include an estimation of the early gluon population with estimates
for the effective inmedium masses of the gluons and the evolution of 
these quantities with time. Work in this direction is currently in 
progress \cite{maj04}.

\begin{acknowledgments}
The authors wish to thank Y. Aghababaie, S. Jeon, J. I. Kapusta, V. Koch, 
C. S. Lam, and G. D. Moore, for helpful discussions. 
This work was supported in part by the Natural Sciences and Engineering Research
Council of Canada, in part by the Fonds Nature et Technologies of Quebec, and in part 
by the Director, Office of Science, Office of High Energy and Nuclear Physics, 
Division of Nuclear Physics, and by the Office of Basic Energy
Sciences, Division of Nuclear Sciences, of the U.S. Department of Energy 
under Contract No. DE-AC03-76SF00098. 
\end{acknowledgments}

\appendix
\section{Contour integration of $T^{\mu \nu \rho}$}

In this appendix, we outline the formal calculation of the 
two gluon photon vertex in the imaginary time formalism, using
the method of contour integration. 
In this case, the standard method of contour integration will 
be modified to allow for the appearance of expressions which 
may be easily generalized from the case at zero density. 
This procedure allows for the construction of the spectator interpretation. 
As mentioned before in Eqs. (\ref{vert1}) and (\ref{vert2}) 
the Feynman rules for the two-gluon-photon 
vertex in the imaginary time formalism are,

\bea 
\mt^{\mu \nu \rho} &=& \frac{-1}{\B} \int \frac{d^3 q}{ (2 \pi)^3 } \sum_n \mbox{Tr} \llb i e \kd_{ki} 
\g^\mu  \frac{i ( \f\fq + m ) }{ \fq^2 - m^2 } i g t^b_{ij} \g^\nu \nn \\
&\times& \frac{i ( \f\fq - \f\fk + m ) }{ (\fq-\fk)^2 - m^2 } i g t^c_{jk} \g^\ro 
\frac{ i ( \f\fq - \f\fp + m ) }{ (\fq-\fp)^2 - m^2 }  \lrb 
\label{vert1a} 
\eea

\bea 
\mt^{\mu \ro \nu} &=& \frac{-1}{\B} \int \frac{d^3 q}{ (2 \pi)^3 } \sum_n \mbox{Tr} \llb i e \kd_{ik} 
\g^\mu  \frac{i ( \f\fq + \f\fp + m ) }{ (\fq+\fp)^2 - m^2 } i g t^c_{kj} \g^\ro \nn \\
&\times& \frac{i ( \f\fq + \f\fk + m ) }{ (\fq+\fk)^2 - m^2 } i g t^b_{ji} \g^\nu 
\frac{ i ( \f\fq + m ) }{ \fq^2 - m^2 }  \lrb  
\label{vert2a}
\eea

\nt Where the trace is implied over both colour and spin indices. 
The zeroth components of each four-momentum are 

\[
q^0 = i(2n+1)\pi T + \mu , \mbox{\hspace{1cm}} p^0 = i2m \pi T, \mbox{\hspace{1cm}} k^0 = i 2j \pi T . 
\]

\nt where $n, m, j$ are integers, $\mu$  is the quark chemical potential.  
 The overall minus sign is 
due to the fermion loop. The sum over $n$ runs over all integers from $-\infty$ to $+\infty$.
As mentioned previously, the momentum dependent and mass dependent parts of the numerators of Eqs. 
(\ref{vert1a}) and (\ref{vert2a}) may be separated as

\bea
\mt^{\mu \nu \rho} &=& \mb^{\mu \A \nu \B \ro \g} {\mathcal{T}_{1}}_{\A \B \g } 
+ \mathcal{A}_1^{\mu \nu \ro}
= \frac{ e g^2 \kd^{bc} }{2 \B} \int \frac{ d^3 q }{ (2 \pi)^3 } \sum_n \mbox{Tr}  \nn \\
& & \Bigg[ \frac{ \mb^{\mu \A \nu \B  \ro \g} q_\A ( q - k )_\B ( q - p )_\g  }
{ ( \fq^2 - m^2 )  ( (\fq - \fk)^2 - m^2 )  ( (\fq - \fp)^2 - m^2 ) } \nn \\
&+& m^2 
\frac{ \ma^{\mu \A \nu \ro} q_\A + \ma^{\mu \nu \B \ro} ( q - k )_\B + \ma^{\mu \nu \ro \g} ( q - p )_\g}
{  ( \fq^2 - m^2 )  ( ( \fq - \fk )^2 - m^2 )  ( ( \fq - \fp )^2 - m^2 )  } \Bigg]
\eea

\bea
\mt^{\mu \ro \nu} &=& \mb^{\mu \A \ro \B \nu \g} {\mathcal{T}_{2}}_{\A \B \g } + \mathcal{A}_2^{\mu \nu \ro}
= \frac{ e g^2 \kd^{bc} }{2 \B} \int \frac{ d^3 q }{ (2 \pi)^3 } \sum_n \mbox{Tr}  \nn \\
& & \Bigg[ \frac{ \mb^{\mu \A \ro \B  \nu \g} ( q + p )_\A ( q + k )_\B ( q )_\g  }
{ ( \fq^2 - m^2 )  ( (\fq + \fk)^2 - m^2 )  ( (\fq + \fp)^2 - m^2 ) } \nn \\
&+& m^2 
\frac{ \ma^{\mu \A \ro \nu} ( q + p)_\A + \ma^{\mu \ro \B \nu} ( q + k )_\B + \ma^{\mu \ro \nu \g} q_\g }
{  ( \fq^2 - m^2 )  ( ( \fq + \fk )^2 - m^2 )  ( ( \fq + \fp )^2 - m^2 )  } \Bigg]
\eea

\begin{figure}[!htb]
  \begin{center}
  \resizebox{5in}{5in}{\includegraphics[0in,1in][9in,10in]{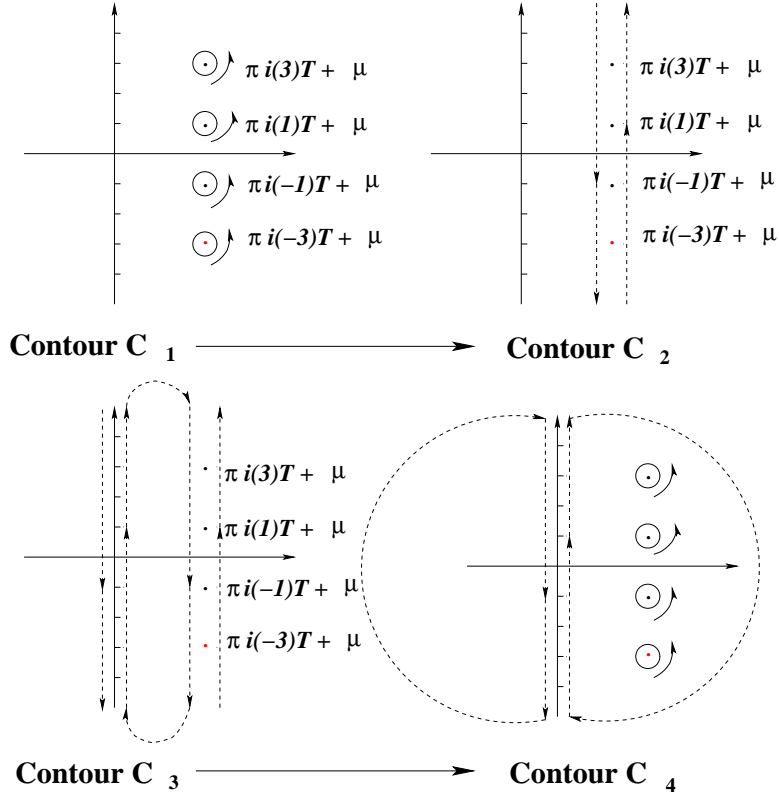}}    
\vspace{-2cm}
    \caption[The contours used to evaluate the Matsubara sum with a finite chemical potential]
{ The contours used to evaluate the Matsubara sum with a finite chemical potential. See text for
details. }
    \label{conts}
  \end{center}
\vspace{-1cm}
\end{figure}

\nt Where $\ma^{\mu \nu \ro \g}$ represents the trace of four $\g$ matrices and 
$\mb^{\mu \A \nu \B \ro \g}$ represents the trace of six $\g$ matrices. 
Employing the methods of residue calculus, the sum over 
$n$ may be formally rewritten as a contour integration over the infinite set of contours 
$ C_1 $ (See Fig. \ref{conts})

\begin{equation}
T \sum_n f(q^0 = i(2n+1)\pi T + \mu ) = \frac{ T }{ 2 \pi i } \oint_{ C_1 }  d q^0  f(q^0) \af 
\B \tanh \left( \af \B ( q^0 - \mu ) \right)  
\end{equation}

The contours $C_1$ may be deformed to those of $C_2$ 
(see Fig. \ref{conts}). These are a set of 
two linear contours meeting at $\pm i \ini$, one from 
$q^0 = -i \ini + \mu + \e \ra q^0 = i \ini + \mu + \e$, 
and another from $i \ini + \mu - \e \ra -i \ini + \mu - \e$. Here, and 
henceforth in all discussions of 
contours, residues and analytic continuations,  
$\e$ will represent a  vanishingly small quantity. One may now proceed by the 
standard method of \cite{kap89} and 
separate a vacuum part, thermal part and a pure density contribution. Instead, 
another set of contours is introduced: 
these run along the $y$ axis from $q^0 =  -i \ini + \e \ra q^0 =  i \ini + \e $ 
and from $ i \ini - \e \ra -i \ini - \e $. 
Admittedly, as $\e \ra 0$ this contour will produce a vanishing contribution. The 
integrand in Eq. (\ref{vert1a}) has six powers of $q^0$ in the 
denominator and only three in the numerator.
Hence it vanishes faster than a linear term as $q^0 \ra \ini$. 
As a result, this quantity obeys Jordan's Lemma and 
the two integration contours around $0$ and $\mu$ may be connected 
by line segments at $\pm i \ini$. These 
line segments shown as curved lines in the third contour of Fig. \ref{conts} 
will have zero contribution to 
the entire integral. The total contour thus obtained is referred to as $C_3$. 
We now split the integrand 
into two, one piece for all the contours on the positive of the $x$ axis 
denoted as $C_3^a$, one piece for the 
sole contour on the negative side of the $x$ axis denoted as $C_3^b$, i.e.,

\bea
\frac{ T }{ 2 \pi i } \oint_{ C_1 }  d q^0 f(q^0) \af \B \tanh \left( \af \B ( q^0 - \mu ) \right) 
&=& \frac{1}{ 2 \pi i }  \int_{ {i \ini - \e}_{C_3^b} }^{ -i \ini - \e } 
d q^0 f( q^0 ) \left( -\af  + \frac{1}{ e^{ \B ( \mu - q^0 )  }  +  1 }  \right)  \nn \\
\mbox{} + \frac{1}{ 2 \pi i }  \left(  \int_{ -i \ini + \e }^{ i \ini + \e } + 
\int_{ i \ini + \mu - \e }^{ -i \ini + \mu - \e } + \int_{ - i \ini + \mu + \e }^{ i \ini + \mu + \e }
\right)_{C_3^a} & &  \!\!\!\!\!\!\! d q^0 f( q^0 ) \left( \af  - \frac{1}{ e^{ \B ( q^0 - \mu )  }  +  1 } 
\right)  .
\eea

The terms may now be separated into a vacuum piece and a matter piece, 
note the similarity between this and 
the zero density separation. In this procedure, we differ from the 
standard method \cite{kap89} in not 
extracting an explicit finite density piece. The main reason for 
the extra contour deformation is
to obtain the final answer in a form where the 
zero density contribution is obvious.
In this spirit, we now reverse the direction of integration in $C_3^b$ 
and note that the vacuum piece
has no poles at $i(2n+1)\pi T + \mu$. Thus the contours in the vacuum 
term may be allowed to overlap 
by setting $\e=0$. We obtain

\bea 
& & \frac{ T }{ 2 \pi i } \oint_{ C_1 }  d q^0 f(q^0) \af \B \tanh \left( \af \B ( q^0 - \mu ) \right)
= \frac{1}{ 2 \pi i }  \int_{ -i \ini }^{ i \ini }  d q^0 f( q^0 ) \nn \\
&+&  \frac{1}{ 2 \pi i }  \int_{ -i \ini - \e}^{ i \ini - \e } 
d q^0 f( q^0 )  \frac{-1}{ e^{ \B ( -q^0 + \mu )  }  +  1 }  \nn \\
&+& \left(  \int_{ -i \ini + \e }^{ i \ini + \e } + 
\int_{ i \ini + \mu - \e }^{ -i \ini + \mu - \e } + \int_{ - i \ini + \mu + \e }^{ i \ini + \mu + \e }
\right)  d q^0 f( q^0 ) \frac{-1}{ e^{ \B ( q^0 - \mu )  }  +  1 }  .
\eea

\nt We now let $\e \ra 0$ on the contours on the positive side 
of the $x$ axis. This procedure will 
deform the two linear contours at $\mu \pm \e$ back to 
the small circles around the points
$ i (2n+1) \pi T $, this part will become similar to the 
initial contour $C_1$. The rest of the contour 
can be closed by including the infinite arc in the $ q^0 = + \ini $ 
direction in the clockwise sense. This multiply 
connected contour is indicated as $C_4^a - C_1$ and displayed on the right of the fourth plot 
in Fig. \ref{conts}. The linear 
contour on the negative side may be closed off as always by the infinite arc 
extending to $ q^0 = - \ini $. This is indicated as $C_4^b$ and 
shown as the left contour in the fourth plot of the figure. 
The contour integration over either contour may be replaced by the sum 
over all the residues at all the poles enclosed by the contour. 
Note that the poles at $i(2n+1)\pi T + \mu$, excluded
by the multiply connected contour, are not to be included in the 
sum over residues.  Thus our final, formal result is,

\bea 
\!\!\!\!\!\!\!\!\!\!\! \frac{ T }{ 2 \pi i } \oint_{ C_1 }  d q^0 f(q^0) 
\af \B \tanh \left( \af \B ( q^0 - \mu ) \right)
&=& \frac{1}{ 2 \pi i }  \int_{ -i \ini }^{ i \ini }  d q^0 f( q^0 ) \nn \\
- \sum_{i} & & \!\!\!\!\! \h(-\w_i) \res[ f( q^0 ) ]  
\frac{1}{ e^{ \B ( -q^0 + \mu )  }  +  1 } \Bigg|_{q^0=\w_i} \nn \\
+ \sum_{i} & & \!\!\!\!\! \h(\w_i) \res[ f( q^0 ) ] 
\frac{1}{ e^{ \B ( q^0 - \mu )  }  +  1 } \Bigg|_{q^0=\w_i} .
\eea

\nt We may substitute the full integrand in Eq. (\ref{vert1a}) to obtain the 
result of contour integration as 

\bea
\mt^{\mu \nu \rho} &=&  \frac{1}{ 2 \pi i }  \int_{ -i \ini }^{ i \ini }  d q^0\int \frac{ d^3 q }{ (2 \pi)^3 }  
\Bigg[ \frac{ e g^2 \kd^{bc} }{2 \B} \Bigg(
\frac{ \mb^{\mu \nu \ro}_{\A \B \g } q^\A ( q - k )^\B ( q - p )^\g  }
{ ( \fq^2 - m^2 )  ( (\fq - \fk)^2 - m^2 )  ( (\fq - \fp)^2 - m^2 ) } \nn \\
&+& 4m^2 \frac{ \gmn ( q - p - k )^\ro  + \gmr ( q - k + p )^\nu  +  \gnr ( q + k - p )^\mu }
{  ( \fq^2 - m^2 )  ( ( \fq - \fk )^2 - m^2 )  ( ( \fq - \fp )^2 - m^2 )  }  \Bigg) \nn \\
&+& \sum_{i} \Bigg\{ \llb \frac{\h(\w_i)}{ e^{ \B ( q^0 - \mu )  }  +  1 } 
- \frac{\h(-\w_i)}{ e^{ \B ( -q^0 + \mu )  }  +  1 } \lrb  \nn \\
&\times& \frac{ e g^2 \kd^{bc} }{2 \B}   \res   
\Bigg( \frac{ \mb^{\mu \nu \ro}_{\A \B \g } q^\A ( q - k )^\B ( q - p )^\g  }
{ ( \fq^2 - m^2 )  ( (\fq - \fk)^2 - m^2 )  ( (\fq - \fp)^2 - m^2 ) } \nn \\
&+& 4m^2 \frac{ \gmn ( q - p - k )^\ro  + \gmr ( q - k + p )^\nu  +  \gnr ( q + k - p )^\mu }
{  ( \fq^2 - m^2 )  ( ( \fq - \fk )^2 - m^2 )  ( ( \fq - \fp )^2 - m^2 )  } \Bigg) \Bigg\}_{q^0=\w_i} \Bigg] .
\label{vert1,2}
\eea

A similar contour analysis as above may be performed for Eq. (\ref{vert2a}), 
with the added extra step 
of setting $q^0 \ra -q^0, \vq \ra -\vq$. This procedure will produce a 
final contour of integration 
which is a mirror image of $C_4$. There will, once again, be an 
infinite semi-circle extending to $+ \ini$ 
connected with the line running from $-i\ini + \e \ra i \ini + \e$. 
There will also be an 
infinite semi-circle extending to $- \ini$ connected to the vertical line 
running on the negative side 
of the $x$ axis. This contour will however be multiply connected 
with the poles at $ -i(2n+1)\pi T - \mu $
excluded from the region bounded by the infinite semi-circle. As before 
these poles shall be excluded 
from the sum over residues. Following this procedure, one obtains the 
result of the contour integration for Eq. (\ref{vert2a}) as,

\bea
\mt^{\mu \ro \nu} &=& - \frac{1}{ 2 \pi i } \int_{ -i \ini }^{ i \ini }  
d q^0 \int \frac{ d^3 q }{ (2 \pi)^3 }  
\Bigg[ \frac{ e g^2 \kd^{bc} }{2 \B} \Bigg(
\frac{ \mb^{\mu \nu \ro}_{\A \B \g } q^\A ( q - k )^\B ( q - p )^\g  }
{ ( \fq^2 - m^2 )  ( (\fq - \fk)^2 - m^2 )  ( (\fq - \fp)^2 - m^2 ) } \nn \\
&+& 4m^2 \frac{ \gmn ( q - p - k )^\ro  + \gmr ( q - k + p )^\nu  +  \gnr ( q + k - p )^\mu }
{  ( \fq^2 - m^2 )  ( ( \fq - \fk )^2 - m^2 )  ( ( \fq - \fp )^2 - m^2 )  }  \Bigg) \nn \\
&+& \sum_{i} \Bigg\{ \llb \frac{\h(\w_i)}{ e^{ \B ( q^0 + \mu )  }  +  1 } 
- \frac{\h(-\w_i)}{ e^{ \B ( -q^0 - \mu )  }  +  1 } \lrb \nn \\
&\times& \frac{ e g^2 \kd^{bc} }{2 \B}   \res 
\Bigg( \frac{ \mb^{\mu \nu \ro}_{\A \B \g } q^\A ( q - k )^\B ( q - p )^\g  }
{ ( \fq^2 - m^2 )  ( (\fq - \fk)^2 - m^2 )  ( (\fq - \fp)^2 - m^2 ) } \nn \\
&+& 4m^2 \frac{ \gmn ( q - p - k )^\ro  + \gmr ( q - k + p )^\nu  +  \gnr ( q + k - p )^\mu }
{  ( \fq^2 - m^2 )  ( ( \fq - \fk )^2 - m^2 )  ( ( \fq - \fp )^2 - m^2 )  } \Bigg) \Bigg\}_{q^0=\w_i}   \Bigg] .
\label{vert2,2}
\eea
 
\nt Note, that the vacuum term is at least naively linearly divergent and thus the 
shift in momentum integrations may not be performed as above. However, 
from Furry's theorem, one obtains that the sum of the vacuum terms from Eqs. (\ref{vert1,2}) 
and (\ref{vert2,2}) must be identically zero. Also note that the presence of the 
thermal distribution functions over quark momenta renders these integrals ultra-violet finite. 
Quark momentum shifts are thus definitely allowed for the thermal 
parts of Eqs. (\ref{vert1,2}) and (\ref{vert2,2}).
Hence, We ignore the vacuum pieces and 
combine the matter pieces of both terms to obtain 
$T^{\mu \nu \rho} = \mt^{\mu \nu \ro} + \mt^{\mu \ro \nu}$ as,

\bea
T^{\mu \nu \rho} &=& \int \frac{ d^3 q }{ (2 \pi)^3 } \sum_{i} \llb \h(\w_i) 
\left( \frac{1}{ e^{ \B ( q^0 - \mu )  }  +  1 } - \frac{1}{ e^{ \B ( q^0 + \mu )  }  +  1 } \right) 
+ \h(-\w_i) \left( \frac{1}{ e^{ \B ( -q^0 - \mu )  }  +  1 } - 
\frac{1}{ e^{ \B ( -q^0 + \mu )  }  +  1 } \right)
\lrb  \nn \\
&\times& \frac{ e g^2 \kd^{bc} }{2 \B}   \res   
\Bigg( \frac{ \mb^{\mu \nu \ro}_{\A \B \g } q^\A ( q - k )^\B ( q - p )^\g  }
{ ( \fq^2 - m^2 )  ( (\fq - \fk)^2 - m^2 )  ( (\fq - \fp)^2 - m^2 ) } \nn \\
&+& 4m^2 \frac{ \gmn ( q - p - k )^\ro  + \gmr ( q - k + p )^\nu  +  \gnr ( q + k - p )^\mu }
{  ( \fq^2 - m^2 )  ( ( \fq - \fk )^2 - m^2 )  ( ( \fq - \fp )^2 - m^2 )  } \Bigg) \Bigg|_{q^0=\w_i}
\eea

\end{document}